\documentclass[12pt]{spieman}  % 12pt font required by SPIE;
\usepackage{amsmath,amsfonts,amssymb}
\usepackage{graphicx}
\usepackage{setspace}
\usepackage{tocloft}
\usepackage{caption}
\usepackage{subcaption}
\usepackage{siunitx}
\usepackage[T1]{fontenc}
\usepackage{enumitem}
\usepackage{microtype}
\usepackage{tabularx}
\usepackage{tikz}
\usetikzlibrary{fit}
\usepackage{float}
\hypersetup{
  pdftitle={Comparative analysis of fiber Bragg grating filter losses inscribed by continuous wave UV and femtosecond-IR lasers for astrophotonics},
  pdfauthor={Aashia Rahman, Ria G. Krämer, Abani Shankar Nayak, Julius Göhring, Piyamas Choochalerm, Anna Maria Weiß, Samuel L. Döpfner, Tim Schleifer, Kalaga Madhav, Martin M. Roth, Stefan Nolte},
  pdfsubject={Astrophotonics, FBG filters},
  pdfkeywords={fiber Bragg grating, astrophotonics, OH filters, insertion loss}
}
\title{Comparative analysis of fiber Bragg grating filter losses inscribed by continuous wave UV and femtosecond-IR lasers for astrophotonics}

\author[a,*]{Aashia Rahman}
\author[b]{Ria G. Krämer}
\author[a]{Abani Shankar Nayak}
\author[a]{Julius Göhring}
\author[c]{Piyamas Choochalerm}
\author[a,d]{Anna Maria Weiß}
\author[b]{Samuel L. Döpfner}
\author[a,e]{Tim Schleifer}
\author[a,d]{Kalaga Madhav}
\author[a]{Martin M. Roth}
\author[b,f]{Stefan Nolte}

\affil[a]{Astrophotonics (innoFSPEC), Leibniz Institute for Astrophysics Potsdam (AIP), An der Sternwarte, 14482 Potsdam, Germany}
\affil[b]{Friedrich Schiller University Jena, Abbe Center of Photonics, Institute of Applied Physics, Albert-Einstein-Str. 15, 07745 Jena, Germany}
\affil[c]{National Astronomical Research Institute of Thailand (NARIT), Don Kaeo, Mae Rim District, Chiang Mai 50180, Thailand}
\affil[d]{Institute of Physics and Astronomy, University of Potsdam, Karl-Liebknecht-Straße 24-25, 14476 Potsdam, Germany}
\affil[e]{Technische Universität Berlin, Straße des 17. Juni 135, 10623 Berlin, Germany}
\affil[f]{Fraunhofer Institute for Applied Optics and Precision Engineering IOF, Center of Excellence in Photonics, Albert-Einstein-Str. 7, 07745 Jena, Germany}

\cftpagenumbersoff{Fig.}
\cftpagenumbersoff{table} 
\begin{document} 
\maketitle

\begin{abstract}
Fiber Bragg grating (FBG) filters have been demonstrated as promising components in astrophotonic instrumentation for near-infrared ground-based observations. Given the photon-starved nature of astronomical applications, it is critical to minimize insertion losses across astrophotonic components. In addition to the insertion loss (IL) introduced by specialty fibers and inscription techniques, FBGs exhibit cladding mode (CM) losses.
In this work, we studied the loss characteristics of five filter lines in three photosensitive fibers, i.e., a low-numerical-aperture (NA) fiber, a high-NA bend-insensitive fiber, and a cladding-mode-suppressed (CMS) fiber, and in a non-photosensitive fiber, SMF-28. The filters were inscribed using two phase mask-based illumination methods: a continuous wave ultraviolet (UV) laser with a complex phase mask allowing for multi-channel filters, and a femtosecond infrared (fs-IR) laser with phase mask integrated shaping apertures for spectral profile control. 
Our results show that UV-inscribed gratings in high-NA bend-insensitive fiber yield the lowest CM losses ($\approx$ \SI{0.5}{dB}) among photosensitive fibers, but exhibit the highest IL (\SI{4.6}{dB}), and FBGs in non-photosensitive SMF-28 fiber, inscribed with fs-IR, achieve the lowest IL (<\SI{0.05}{dB}) with a comparatively higher CM loss (\SI{0.93}{dB}).
To reduce the high IL in high-NA fiber, we explored tapering and bridging methods and report that bridging reduces IL by $\sim \SI{3}{dB}$. We show that both filter platforms remain viable for integration into an astrophotonic system, with IL below \SI{1}{dB}. Finally, we propose a compact bridge-fiber scheme with the potential to further reduce IL to below \SI{0.5}{dB} while reducing the number of bridging fibers and, consequently, the number of splice junctions by 50\%.

\end{abstract}


{\noindent \footnotesize\textbf{*}Aashia Rahman,  \linkable{arahman@aip.de} }
\begin{spacing}{1}  

\section{Introduction}

The study of the star formation history from the early universe to cosmic noon
\cite{Steidel_1999}, in particular in view of the surprising discovery of a grand-design spiral galaxy at redshift $z \gtrsim 4$ \cite{Jain_2025}, or of galaxies at
ultra-high redshift $9 < z < 12$ \cite{Adams_2023}, made possible with the James
Webb Space Telescope (JWST), illustrates the need in contemporary astrophysics for
spectroscopy in the near infrared (NIR). However, even the light-collecting power of
JWST is limiting the ability to measure metallicities to the analysis of nebular
emission line spectra \cite{Sanders_2024}. The study of stellar populations and their
kinematics through absorption lines, as exercised in the local universe, will at best be
reserved for low/medium spectral resolution spectroscopy with the upcoming generation
of ground-based extremely large telescopes such as the ELT, GMT, or TMT. However, NIR spectroscopy at these facilities will be hampered by atmospheric hydroxyl (OH)
emission lines \cite{Maihara_1993, Meinel}, which are unfortunately numerous and bright in the $J$ and $H$ bands
where the rest-frame diagnostic optical lines are found for redshift $z \gtrsim 2$
galaxies. For faint fluxes of galaxies at high redshift, a high spectral resolution
for OH avoidance is not an option. At low and medium spectral resolution, however, the
extended scattered light wings of OH emission lines fill the gaps of the interline
continuum, therefore severely limiting the ability for faint continuum spectroscopy
needed for stellar population analysis in galaxies at high redshift. Although at the OH
emission line wavelengths proper and within their natural bandwidth, the detection of
light from faint astronomical objects is hopeless, one must avoid the broad
emission line wings spoiling the interline continuum. To resolve this dilemma, one is forced to filter bright OH lines with a very narrow bandwidth before the light is even allowed to enter the optical system; as reviewed by Ellis \& Bland-Hawthorn \cite{ellis2008case}, several attempts to filter out OH lines downstream within a spectrograph have failed.

The pre-spectrograph filtering capability can be achieved effectively using multi-notch filters based on fiber Bragg gratings (FBGs) \cite{ellis2008case}, integrated with photonic lanterns (PLs) \cite{Joss:2011}. The first complete set of 105 multi-notch FBG filters demonstrated for ground-based NIR astronomy (\SI{1466}{nm} to \SI{1700}{nm}) was fabricated for GNOSIS \cite{Trinh_2013} and its successor, PRAXIS \cite{Ellis2020}. The demonstrations of GNOSIS/PRAXIS relied on advanced mathematical modeling \cite{SKAAR2001, SKAAR2002, Buryak:03, Bland-Hawthorn:08} to design the multi-notch filters and highly specialized ultraviolet (UV) illumination-based inscription techniques \cite{Liu, Petermann2002, Stepanov} for their fabrication. Moreover, a custom UV-photosensitive fiber, CMS8 (Nufern/Coherent), was used to support the inscription of low-loss filter lines and enable compatibility with standard fibers in astronomical instrumentation. As CMS8 is no longer commercially available, identifying alternative fiber platforms that can support low-loss OH-filter fabrication, efficient coupling to standard fibers, and compatibility with downstream astronomical instrumentation has become a critical requirement.

Apart from intrinsic scattering and absorption losses that occur in FBG fabrication, two loss mechanisms are particularly critical for OH-filters: cladding mode (CM) losses and insertion loss (IL). CM losses originate from counter-propagating cladding modes that are excited by refractive-index perturbations in the fiber core introduced during the FBG inscription process \cite{Mizrahi, Ivanov_2006, Li_2020}. These CM resonances appear as a series of dips in the short-wavelength part of the FBG transmission spectrum; an effective suppression of the CM resonances is a key requirement for astronomical FBG filters \cite{Jovanovic_2023}, which can otherwise introduce unwanted spectral features in the filter's response.  The IL associated with UV photosensitive fibers arises mainly from mismatches in the numerical aperture (NA) between these fibers and standard fibers, and from fiber attenuation due to the high dopant concentrations in the core of these photosensitive fibers. Hydrogen loading enables the inscription of FBGs in non-photosensitive fibers using UV-illumination; however, this approach introduces additional complexity, including stringent requirements on hydrogenation control, wavelength accuracy\cite{Gbadebo2018}, post-annealing, and long-term stabilization\cite{Grobnic}. An alternative approach to inscribing FBGs in non-photosensitive fibers is using a femtosecond laser. In recent years, Goebel et al. \cite{Goebel:18} have shown that direct femtosecond infrared (fs-IR) writing provides a promising alternative route for fabricating multi-notch OH filters. This approach offers greater flexibility in fiber choice and thus potentially enables reduced IL. Nevertheless, CM losses are intrinsic to FBG-based filters, irrespective of illumination methods. 
As we look ahead to the era of extremely large NIR telescopes, the development of low-loss FBG-based filters will be essential to fully exploit the scientific capabilities of these facilities. These requirements are further amplified by the development of next-generation miniaturized astronomical instrumentation incorporating photonic integrated circuits (PICs), which places additional demands on filter integration with the PICs. The field of astrophotonics \cite{Bland-Hawthorn:09} has advanced rapidly in the past two decades \cite{Dinkelaker:21, Roth:2023, Norris:24, Ellis:2024}, with instruments such as GRAVITY \cite{GRAVITY} on the ESO’s Very Large Telescope (VLT), serving as a compelling testament to the potential of PICs in astronomy. However, to build an astrophotonic facility instrument, efficient injection of telescope light into photonic devices would need further development in several key areas, as addressed in the roadmap by Jovanovic et al. \cite{Jovanovic_2023}, including integration of OH-filters within the astrophotonic system. Therefore, identifying an OH-filter fiber platform with reduced IL and CM losses that is compatible with fiber-to-PIC integration is essential for future astrophotonic instrumentation.

To our knowledge, this study provides the first detailed comparison of IL and CM losses in filter lines across different fibers under UV- and fs-IR-illumination for astronomical applications. Instead of implementing the full set of $\sim$ 105 OH-suppression lines (in the $H$ band), we evaluated five closely spaced representative filter lines as a controlled test case to compare fiber- and inscription-dependent loss mechanisms. The paper is organized as follows: first, we assess the IL and CM losses of filters within stand-alone fiber platforms. Here, we describe the fabrication and loss characterization methods and compare the performance of representative photosensitive and non-photosensitive fibers. Hydrogen-loaded fibers were not considered in the present study due to their complexity in inscription, as previously stated. In the second section, we investigate IL-minimizing approaches, namely, tapering and bridging fiber methods, to improve the performance of UV-inscribed filters in photosensitive fibers. Finally, in the last section, we discuss the overall throughput of such filters relevant to astrophotonics for next-generation miniaturized instrumentation, and propose a suitable fiber configuration scheme for FBG filters inscribed using both UV and fs-IR techniques.

\section{FBG fabrication and filter loss characterization}
\subsection{FBG fabrication}\label{FBG_fab}
We employed phase mask-based inscription techniques to generate the periodic refractive index modification in the fiber core for the FBG with two main illumination methods: UV and fs-IR laser.
The UV-illumination uses an adapted fabrication technique, utilizing a custom-designed complex phase mask \cite{Luo, Luo:25}, which was designed to incorporate flexible partial spatial overlap among the five channels. This allows for the fabrication of multi-channel complex gratings with optimized trade-offs between the interline losses and overall grating length. As the phase mask was designed to demonstrate the concept and fabrication process of this design approach, additional complexities, such as apodization, were not incorporated at this stage.
For the fs-IR phase mask scanning technique \cite{Thomas2012}, a different approach was employed. Here, a shaping aperture \cite{Kramer:25} was directly implemented into the phase mask to tailor the profile of the induced refractive index modification, enabling reproducible control over the shape of the spectral response for the precise fabrication of OH-filters.

The two illumination techniques differ distinctly in their absorption process: in UV-based inscription, the single-photon absorption requires photosensitive material, generally confining the refractive index modification to the photosensitive fiber core. However, the core mode field reaches partially into the fiber cladding enabling coupling to cladding modes and thereby causing CM losses. Therefore, fibers with confined core mode fields or optimized for cladding mode suppression can effectively reduce these losses \cite{Dong:00}. In contrast, fs-IR inscription relies on multi-photon absorption, not limited to photosensitive material. CM losses can be mitigated by expanding the refractive index modification to the cladding and optimizing the scan of the small focal spot to achieve homogeneous coverage \cite{Grobnic:04}.

We characterized IL and CM losses for five filter lines inscribed with UV-illumination in photosensitive fibers: (a) boron doped high photosensitivity with low NA (Fibercore PS1250/1500), (b) high photosensitivity (germania content 5 times to that of an SMF-28) with high NA (Fibercore SM1500(4.2/125)), and (c) cladding-mode-suppressed (Coherent CMS2). For fs-IR inscription, we investigated (a) non-photosensitive fiber (SMF-28) and (b) CMS2 (Coherent CMS2) for the losses. The filters were inscribed in each fiber under controlled and reproducible conditions.
Table \ref{tab:fiber_Para} consolidates the important parameters of these fibers considered in this work.

\begin{table*}[htbp]
\caption{Parameters of the fibers investigated \cite{PS1250, SM1500, CMS2, SMF-28}}
  \centering
\begin{tabular}{|c|c|c|c|c|}
\hline
Parameter (@\SI{1550}{nm})    & PS1250    & SM1500(4.2) & CMS2 & SMF-28\\
         \hline 
           NA              & 0.12 - 0.14  & 0.29 - 0.31 & 0.14 & 0.14\\
           \hline
     MFD ($ \rm{\mu m}$) & 8.8 - 10.6 & 4.0 - 4.5 & 9.6 ± 0.8 & 10.5\\
     \hline
     Attenuation (dB/km)   & 120 & {$\rm{\leq1.5}$} & - & {$\rm{\leq0.2}$} \\
            \hline   
    \end{tabular}
   \label{tab:fiber_Para}
\end{table*}

\begin{sloppypar}
\subsection{IL: Estimation }
 
IL consists of losses due to splicing dissimilar fibers and intrinsic fiber attenuation. The splice-loss can be further broken down into angular mismatch between the splice junctions 
and mode-field diameter (MFD)\-/NA mismatch.
The splice-loss $SL$ is given in decibel, obtained from the power transmission coefficient $T$ by \(SL = -10\log(T)\). $T$ is determined from Marcuse's equation \cite{Marcuse:1977}, 
%Eq.~\ref{eq:Mar}
\end{sloppypar}

\begin{equation}
T = \left( \frac{2 w_1 w_2}{w_1^2 + w_2^2} \right)^2
\exp\!\left[
-\,\frac{2 \left( \pi n_2 w_1 w_2 \theta \right)^2}
{\left( w_1^2 + w_2^2 \right)\lambda^2}
\right]
\label{eq:Mar}
\end{equation}

where $w_1$ and $w_2$ are the mode-field radii of the two fibers, $\theta$ is the angular misalignment, $n_2$ is the refractive index (RI) of the core of the receiving fiber, and $\lambda$ is the wavelength at which the splice-loss is estimated. For calculation, RI $n_2$ is assumed to be 1.45 and $\theta$ is considered to be $\leq1$$^{\circ}$, as we ensured that the cleave angle of all fibers meets the criteria of \(< 0.5^\circ\). The overall IL, combining $SL$ and attenuation, was estimated using the fiber parameters from their corresponding datasheets \cite{PS1250, SM1500, CMS2, SMF-28}, as consolidated in Table \ref{tab:fiber_Para}.

\subsection{IL and CM loss: Measurement}
\begin{figure}[ht]
\centering
\includegraphics[width=0.8\linewidth]{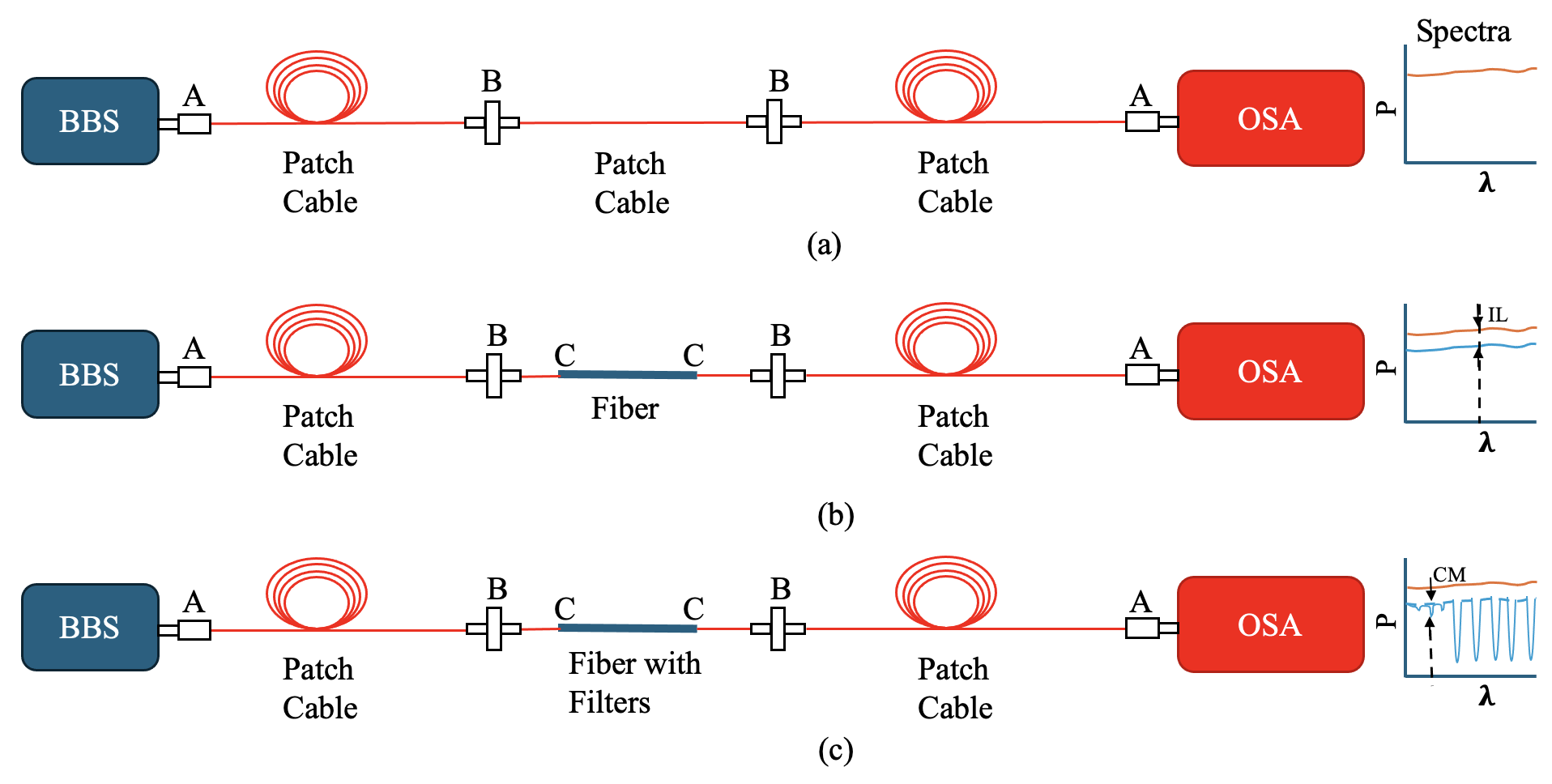}
\caption{Schematic for illustrating the IL and CM measurement setup. (a) The reference spectrum is recorded by connecting the BBS and the OSA. (b) Prior to filter fabrication, IL is measured after the fiber is connected to the BBS and OSA. (c) After all the filters are fabricated, CM is measured at the wavelength where CM loss is highest. 
BBS: Broadband source, OSA: Optical spectrum analyzer, A: FC/PC connector, B: FC/APC-FC/APC adapter, C: spliced junctions between the fiber for filter fabrication and the patch cable.}
\label{fig:Loss_Measurement}
\end{figure}

We measured the IL and CM losses from the experimental data. 
Fig.~\ref{fig:Loss_Measurement} illustrates the schematics of the setup for measuring the IL and CM losses.

In Fig.~\ref{fig:Loss_Measurement}(a), the reference spectra are recorded by connecting a broadband source (BBS) and an optical spectrum analyzer (OSA) through three patch cables in series connected via fiber adapters B. In Fig.~\ref{fig:Loss_Measurement}(b), the center patch cable is cut, and the fiber under investigation is fusion-spliced with the two resulting ends of the cut patch cable, forming the splice junctions C. In this configuration, spectra are recorded and the corresponding IL is measured at \SI{1550}{nm} by comparing the transmission in this setup with the reference transmission from Fig.~\ref{fig:Loss_Measurement}(a). The measured IL consists of splicing losses from angular and MFD mismatches between the fiber and the patch cables at C, along with fiber attenuation due to the fiber length between the C junctions. In the case of fs-IR-inscription in SMF-28 the splice-loss to patch cables can be neglected. The main IL arises from broadband absorption at the refractive index modification; therefore, it is measured at longer wavelengths than the highest resonance wavelength, where no CM coupling occurs, by taking the difference before and after FBG inscription.
Finally, as shown in Fig.~\ref{fig:Loss_Measurement}(c), CM loss is measured at the wavelength where the highest loss is observed (i.e., the largest transmission dip at the shorter wavelength side of the filters), after all five filters have been inscribed.

\subsection{Filter inscription and loss comparison across the fibers}
In this section, we present the filter inscription and compare the IL and CM loss across the fibers mentioned in Section \ref{FBG_fab}. 

For UV inscription, a \SI{500}{mW}, \SI{244}{nm}, frequency-doubled argon ion laser was utilized. To achieve an FBG array of 5 resonances (<\SI{-20}{dB}), a fluence of $\SI{7.87}{kJ/cm^2}$ was optimized for filter fabrication. The total length of the fiber containing the five filters was restricted to $\sim$ \SI{6}{cm}. However, longer fiber lengths ($\sim$ \SI{2.5}{m}) were used to obtain a realistic estimate of IL in a real scenario of multi-notch filters, taking fiber attenuation into account. All the photosensitive fibers were fusion spliced with patch cables (P3-1064Y-FC consisting of fiber type HI1060-J9)\cite{thorlabs_datasheet_2026} at both ends to measure the transmission spectra of the filters. P3-1064Y-FC was used to minimize micro-bend-induced attenuation. For estimating IL, using Eq. \ref{eq:Mar} \cite{Marcuse:1977}, the MFD of HI1060-J9 \cite{thorlabs_datasheet_2026_1} was derived to be $\approx$ \SI{9.5}{\micro\meter} at \SI{1550}{nm}. IL and CM losses were measured using a BBS (Thorlabs ASE730) and an OSA (Yokogawa AQ6375B) with a resolution of \SI{50}{pm}, in a configuration, as shown in Fig.~\ref{fig:Loss_Measurement}.

For fs-IR inscription, an \SI{800}{nm}, \SI{100}{fs}, \SI{1}{kHz} Ti-Sa laser system was utilized. The inscription laser beam was focused with a \SI{25}{mm} cylindrical lens into the fiber core, using pulse energies of \SI{400}{\micro \J} to \SI{475}{\micro \J} and scanning speeds of $10$ to \SI{20}{mm/min}. The phase mask periods range from \SI{1068}{nm} to \SI{1073}{nm}, providing the first order reflection resonance for the targeted OH-lines. To match their spectral bandwidth and increase efficiency, the phase mask has embedded apertures, providing grating length and apodization control. The filters were inscribed subsequently, repeating the inscription scan until the targeted strength (<\SI{-20}{dB}) was reached. Prior to inscription, the SMF-28 was fusion spliced to SMF-28 pigtails, and the transmission spectra were measured using an interrogator (Micron Optics sm125) with a resolution of \SI{5}{pm}. 

\subsubsection{PS1250/1500 and SMF-28}

An FBG array of five resonances (<\SI{-20}{dB}) was inscribed in the
\SIrange{1546}{1552}{\nano\meter} range, targeting the OH-lines at
\SIlist{1546.214;1547.423;1550.088;1550.979;1551.788}{\nano\meter}. 
These five lines were selected from a high-density OH-line region with deep notches in the $H$ band \cite {Trinh_2013}, providing a representative subset for a proof-of-concept characterization of IL and CM losses.

The arrays were inscribed in PS1250/1500 (hereafter PS1250) using UV and in standard
SMF-28 using fs-IR inscription methods.
As UV-inscribed FBG loss studies were restricted to non-hydrogenated fibers, it was essential to use a fiber with high photosensitivity to achieve reflectivities of $\approx$ 99$\%$. To ensure accurate inscription of the gratings at the target OH wavelengths, the stress-optic coefficient of the fiber was experimentally measured and incorporated into the final design \cite{Luo:25}.

\begin{figure}[h]

\begin{subfigure}{0.5\textwidth}
\includegraphics[width=1.0\linewidth]{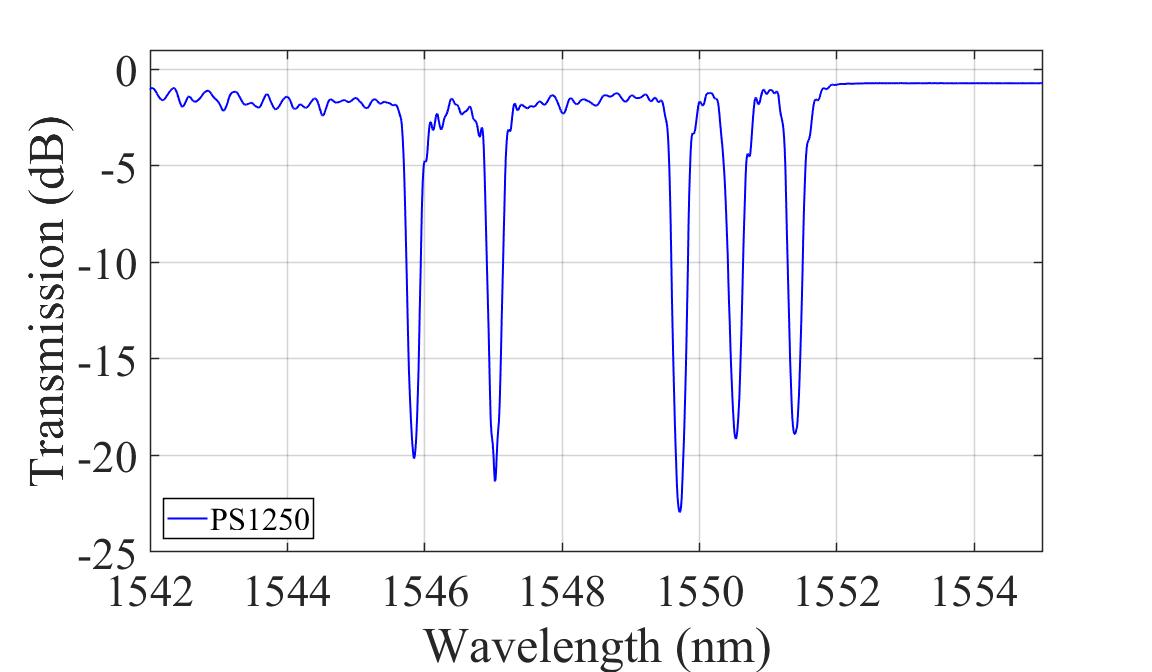} 
\caption{}
\end{subfigure}
\hfill
\begin{subfigure}{0.5\textwidth}
\includegraphics[width=1.0\linewidth]{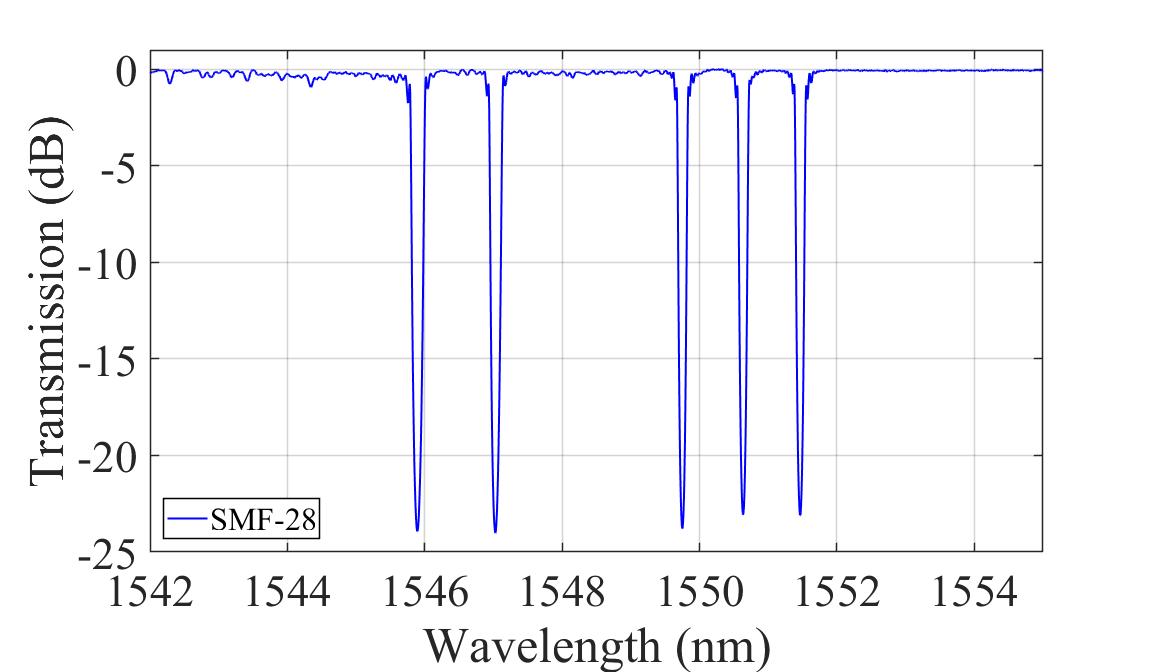}
\caption{}
\end{subfigure}
\caption{Filter transmission spectra. (a) UV-illumination - PS1250 and (b) fs-IR-illumination - SMF-28.
}
\label{fig:PS_SMF}
\end{figure}

 Fig.~\ref{fig:PS_SMF} compares the performance of filter inscriptions in PS1250 and SMF-28 fibers.  
For PS1250, the IL was estimated using Eq.~\ref{eq:Mar}, considering the respective fiber attenuation, and was also measured (as shown in Fig.~\ref{fig:Loss_Measurement}) at \SI{1550}{nm}. The CM losses for PS1250 and SMF-28 were evaluated at the wavelengths corresponding to the maximum CM attenuation. 
For PS1250, the maximum IL estimated at \SI{1550}{nm} is $\leq$\SI{0.85}{dB}, with a maximum angular misalignment of 1$^\circ$; the measured IL was approximately \SI{0.72}{dB}. The maximum CM loss was measured to be \SI{2.39}{dB} at \SI{1544.52}{nm}. For SMF-28, the measured IL was below \SI{0.05}{dB}, and the CM loss was \SI{0.93}{dB} at \SI{1544.07}{nm}.

The fs-inscribed FBG in SMF-28 shows, compared to literature \cite{Grobnic:04}, relatively high CM loss. This is due to the utilized shaping aperture inscription, increasing CM loss due to the inhomogeneous cross-sectional coverage \cite{Kramer:25}. With standard fs-IR-phase mask inscription, CM loss of \SI{0.2}{dB} is achievable. 

PS1250 is a boron-doped photosensitive fiber, and due to boron doping, the attenuation in this fiber is high (\SI{120}{dB/km}). A small fiber length can significantly reduce the IL, for example, a fiber length of about 50 cm will reduce the estimated IL from $\leq$\SI{0.85}{dB} to $\leq$\SI{0.35}{dB}. However, CM losses are very high in this fiber; this is likely attributable to its low NA and correspondingly large MFD. A larger MFD increases the overlap integral between the core and cladding mode fields, thus enhancing grating-induced core-to-cladding mode coupling and resulting in significantly higher CM losses \cite{Dong:00}. To mitigate CM losses for UV-inscription, we next chose two different fibers: a high-photosensitivity fiber (germania content 5 times that of an SMF-28) with high NA (SM1500(4.2/125)) and a fiber with suppressed cladding modes (CMS2).
 
 \subsubsection{SM1500(4.2/125)}\label{SM1500(4.2)}
\begin{figure}[h]
\centering

\begin{subfigure}[t]{0.48\textwidth}
        \centering
        \includegraphics[height=5.0cm]{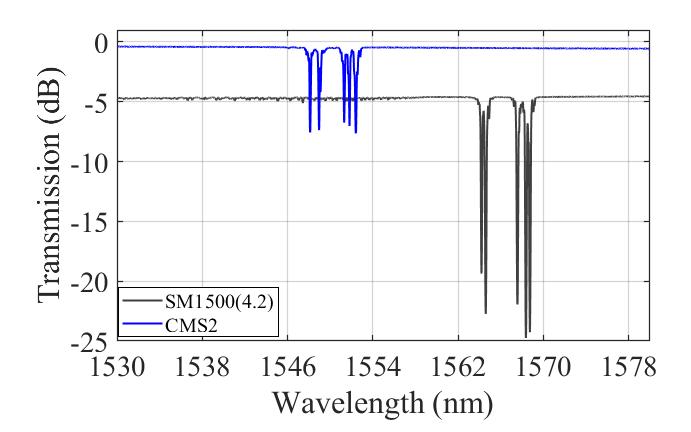}
        \caption{}
        %\label{fig:panel_a}
    \end{subfigure}
    \hfill
    \begin{subfigure}[t]{0.48\textwidth}
        \centering
        \includegraphics[height=5.0cm]{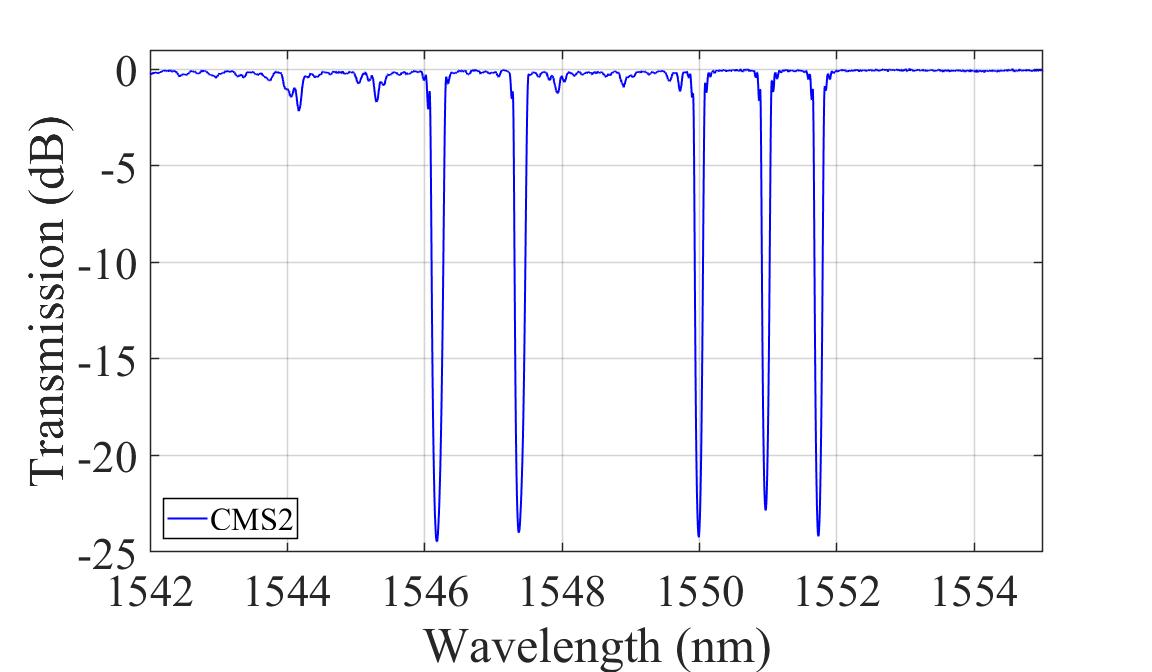}
        \caption{}
        %\label{fig:panel_b}
    \end{subfigure}

\caption{Filter transmission spectra. (a) UV-illumination - SM1500(4.2) and CMS2. Resonance wavelengths of FBGs in SM1500(4.2) and CMS2 are at longer wavelengths due to the higher effective RI of the core mode in those fibers compared to that in PS1250. (b) fs-IR-illumination - CMS2.
}
\label{SM1500}
\end{figure}

 The primary rationale for choosing SM1500(4.2/125) (we will refer to this fiber as SM1500(4.2) hereafter) is its relatively high-NA $\approx$ 0.3, which ensures better mode confinement compared to low-NA (<0.15) photosensitive fibers. Another key advantage is its high bend insensitivity, allowing the fiber to be tightly coiled without significant bend losses. This is particularly useful when increasing the number of notches (especially for OH-filters), as the fiber can be compactly coiled between sections containing filter arrays. Additionally, the attenuation of SM1500(4.2) is about 80 times lower than that of PS1250, making it suitable for use over long fiber lengths. However, due to its small MFD ($\approx$ \SI{4.2}{\micro\meter}), connecting it with a standard patch cable, P3-1064Y-FC, where the MFD is $\approx$ \SI{9.5}{\micro\meter} introduces high IL, i.e., we estimated (Eq.~\ref{eq:Mar}) it to be as high as about \SI{5.5}{dB} at \SI{1550}{nm}, with a maximum angular misalignment of 1$^\circ$. Fig.~\ref{SM1500}(a) shows the spectra (in black) obtained with SM1500(4.2). The measured IL was \SI{4.6}{dB} at \SI{1550}{nm}. However, CM losses are much lower in SM1500(4.2), $\leq$ \SI{0.5}{dB} (measured at \SI{1547.43}{nm}), even lower than the aperture-shaped fs-inscribed filters in SMF-28.

\subsubsection{CMS2}
Coherent has a specialized fiber series (CMS-series) that offers cladding mode suppression and enables tighter channel spacing for various applications in telecommunications. For our work, CMS2 was selected from Coherent’s CMS-series because its MFD closely matches that of standard single-mode fiber patch cables (Table \ref{tab:Allfiber}), which minimizes the splice-loss and therefore IL is reduced. For UV-inscription, although the IL and CM losses are low in CMS2, both around \SI{0.5}{dB}, CMS2 exhibits low photosensitivity, and we were unable to achieve FBG reflectivities exceeding 82\% (Fig.~\ref{SM1500}(a)(spectra in blue)). This fiber would be a suitable choice if used in conjunction with hydrogenation to improve photosensitivity at 244 nm. However, for stronger gratings (reflectivities > 96\%) the CM loss in CMS2 is expected to increase. As the performance of CMS2 fiber cannot be fully explored unless it is hydrogenated, a comparison with other fibers, PS1250, SMF-28 and SM1500(4.2) with stronger FBGs (reflectivities > 96\% ) could not be evaluated. At this stage, we do not pursue further investigation of CMS2 under UV-illumination; this will be addressed in future work focusing on the effects of hydrogenation in filter fabrication for astronomical applications.

We next inscribed filters in CMS2 using fs-IR-illumination. Fig.~\ref{SM1500}(b) presents the filter spectra obtained for CMS2. The filters were inscribed to a reflectivity of $\approx$ 99\%, 
comparable to the SMF-28 fs-IR filters. The filters show very low IL (<\SI{0.05}{dB}). However, a significant CM loss of \SI{2.17}{dB} occurs at \SI{1544.16}{nm}. Possibly, the increased refractive index modification area of the fs-inscription extends into the depressed cladding of CMS2, altering the fields of the core and cladding modes, thereby leading to an increased overlap integral and, consequently, higher CM loss. Therefore, to exploit the cladding mode suppressing nature of the CMS2, the fs-inscription parameters need to be adapted, which, however, is part of future work.

Table \ref{tab:Allfiber} summarizes the results for all investigated fibers. FBG filters inscribed in SMF-28 exhibit the best overall performance among the fibers investigated. However, SM1500(4.2) remains a viable alternative due to its low CM losses, provided that the IL can be further reduced. 
In the literature, different methods are available to mitigate IL due to the dissimilar MFDs of the fibers, either by tapering \cite{Yablon:2005, Miller:18} or using bridging fibers \cite{Holmes1990MatchingFF, Chandan:2004}. In Section \ref{IL_red}, we explore both tapering and bridging options for SM1500(4.2).

\setlength{\tabcolsep}{0.5pt}
\begin{table*}[htbp]
\caption{Loss Comparison - All Investigated Fibers}
  \centering
\begin{tabular}{|c|c|c|c|c|c|}
\hline
Loss    & PS1250    & SM1500(4.2) & CMS2 (UV)\textsuperscript{\#} & SMF-28 & CMS2 (fs-IR) \\
         
         \hline 
          Estimated IL [dB] & {$\rm{\leq0.85}$} & {$\rm{\leq5.5}$} & {$\rm{\leq0.55}$} & - & -\\
          & (@\SI{1550}{nm}) & (@\SI{1550}{nm}) & (@\SI{1550}{nm}) & &\\
          \hline
        Measured IL [dB] & 0.72 & 4.6 & 0.53 & $<$0.05 & $<$0.05\\
         & (@\SI{1550}{nm}) & (@\SI{1550}{nm}) & (@\SI{1550}{nm}) & (@>\SI{1552}{nm}) & (@>\SI{1552}{nm}) \\
         \hline 
        CM loss [dB]& 2.39  & 0.49 %(-5.098) 
        & 0.52 & 0.93 & 2.17\\
            & (@\SI{1544.52}{nm}) & (@\SI{1547.43}{nm}) & (@\SI{1547.07}{nm}) & (@\SI{1544.07}{nm}) & (@\SI{1544.16}{nm})\\
            \hline   
    \end{tabular}
    \vspace{2mm}
\begin{flushleft}
\footnotesize{\textsuperscript{\#} Due to weaker filter generation, the filter-loss performance of non-hydrogenated CMS2(UV) is not directly comparable with PS1250, SM1500(4.2), SMF-28, and CMS2(fs-IR). However, for the comprehensiveness of this study, we still present the data of CMS2(UV) in Table \ref{tab:Allfiber}.}
\end{flushleft}

   \label{tab:Allfiber}
\end{table*}

\section{Methods to reduce IL in SM1500(4.2)}\label{IL_red}

The primary contribution to IL arises from splicing losses caused by MFD mismatch (\SI{4.2}{\micro\meter} of SM1500(4.2)) to standard fibers (\SI{10.4}{\micro\meter} of SMF-28). In the following, two mitigation strategies, adiabatic tapering and bridging fibers, will be evaluated.

\subsection{Adiabatic Tapering}\label{IL_AdbTap}
To reduce the losses arising from splice-loss caused by MFD mismatch between the fibers, a controlled transition of the MFD is required, either by tapering the larger-mode fiber or by increasing the effective MFD of the smaller-mode fiber during splicing via controlled heat treatment, or a combination of both. The approach involving heat treatment during splicing/tapering requires precise control of dopant out-diffusion \cite{Shiraishi:1990, Tam:1991} from the core to the cladding and has therefore not been pursued in this work. 

\begin{figure}[h]

\begin{subfigure}{0.5\textwidth}
\includegraphics[width=1.0\linewidth]{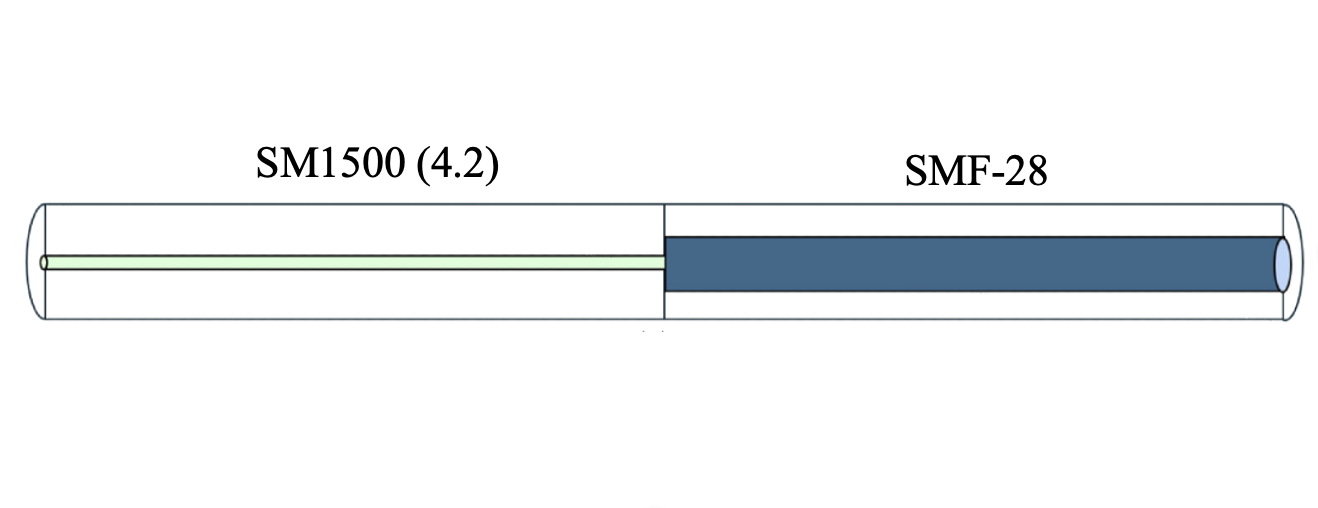} 
\caption{}
\end{subfigure}
\hfill
\begin{subfigure}{0.5\textwidth}
\includegraphics[width=1.0\linewidth]{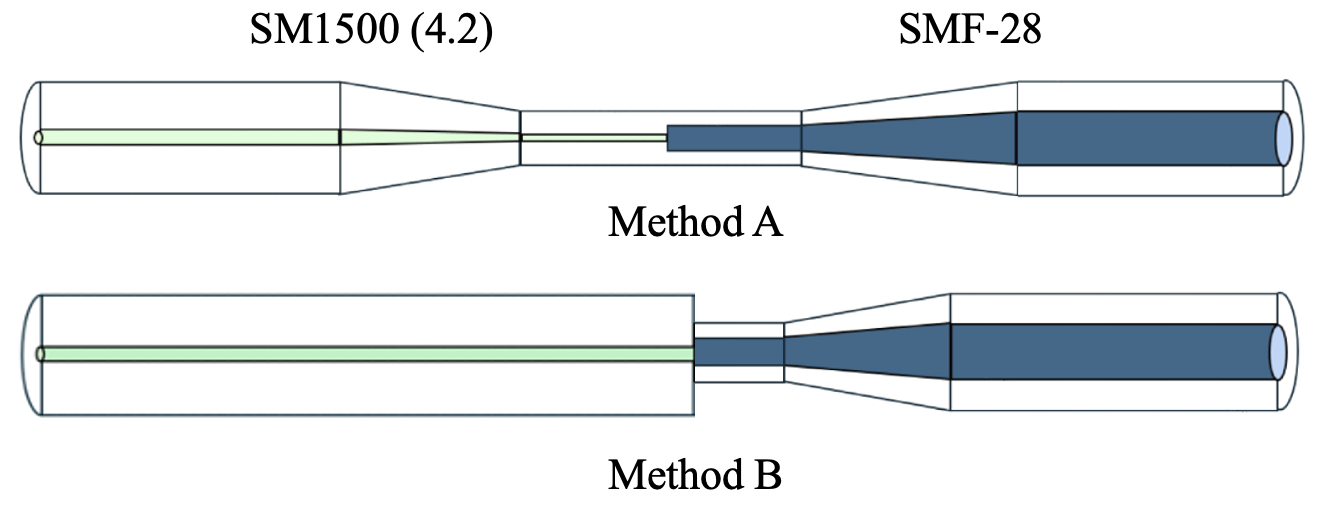}
\caption{}
\end{subfigure}
\caption{Schematic illustrating methods carried out in this work (a) direct fusion splice between SM1500(4.2) and SMF-28. (b) Method A: SM1500(4.2) to SMF-28 are spliced first and then the spliced junction is tapered; Method B: SMF-28 is pre-tapered, cleaved and then spliced with SM1500(4.2).
}
\label{fig:schematic}
\end{figure}

To investigate tapering methods for reducing splice-loss, we examined short sections of SM1500(4.2) (without filters) and standard SMF-28 fiber. Fig.~\ref{fig:schematic} illustrates schematics of direct splicing (Fig.~\ref{fig:schematic}(a)) and tapering (Fig.~\ref{fig:schematic}(b)) methods between SM1500(4.2) and SMF-28. In Fig.~\ref{fig:schematic}(b), Method A denotes fusion splicing followed by tapering of the splice junction, while Method B denotes pre-tapering of SMF-28 prior to splicing with SM1500(4.2). We used Sumitomo Electric (TYPE-Q101-CA), a high-precision core-alignment fusion splicer for direct splicing, and the Vytran glass processing system (GPX-3000 series) for tapering/pre-tapering. A uniform transition/taper length of \SI{4}{\centi\meter} was maintained across all samples to ensure adiabatic conditions \cite{Davenport:21}, while taper diameters were varied for both Method A and Method B. We identified \SI{60}{\micro\meter} to be the minimum reliable limit for the taper diameter, which is imposed by the combined operational constraints of the Vytran tapering system and the Sumitomo fusion splicer, even under optimized parameters. Consequently, four cladding diameters of \SI{60}{\micro\meter}, \SI{65}{\micro\meter}, \SI{70}{\micro\meter}, and \SI{75}{\micro\meter} were 
chosen as the target taper diameters for the methods A and B.

\begin{figure}[ht]
\centering
\includegraphics[width=0.4\linewidth]{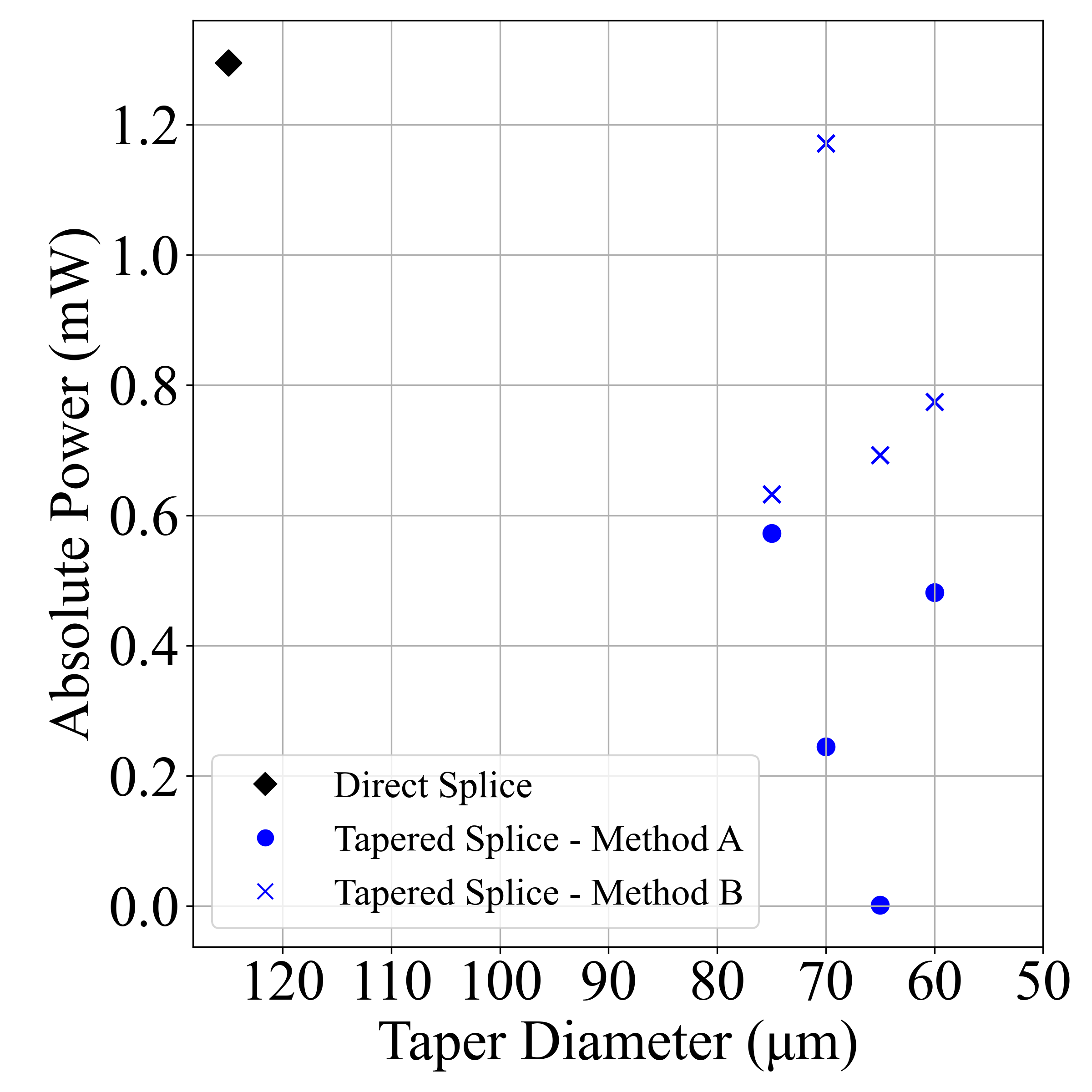}
\caption{Measured transmitted power for direct fusion splice, Method~A, and
Method~B at $\lambda = \SI{1550}{nm}$ (tunable laser, $P = \SI{10}{mW}$). The reported power values represent the transmitted power averaged over the acquisition period. Taper diameter refers to the tapered cladding diameter.}
\label{fig:Results}
\end{figure}
The transmitted power at $\lambda = \SI{1550}{nm}$ for each configuration is
shown in Fig.~\ref{fig:Results}. %Direct splicing yields the highest throughput
%($\sim$\SI{1.3}{\milli\watt}), exceeding both tapered methods: Method~B
%reaches $<$\SI{1.2}{\milli\watt} and Method~A only $<$\SI{0.6}{\milli\watt}.
%Contrary to expectation, reducing the taper diameter further to match the MFD
%of SM1500(4.2) did not improve transmission.
 We observe that (i) direct splicing yields a higher throughput of $\sim$ \SI{1.3}{\milli\watt}, compared to the combination of splicing and tapering (Methods A and B). The maximum throughput was $<$ \SI{0.6}{\milli\watt} in Method A and $<$ \SI{1.2}{\milli\watt} in Method B. 
A possible explanation for this observation is as follows:  The high temperatures during the direct splicing process promote dopant diffusion from the core of the high-NA fiber into the cladding \cite{Shiraishi:1990, Tam:1991, Rahman2010}, thereby decreasing the core RI at the splice, which in turn decreases the fiber NA. This reduced NA in the SM1500(4.2) fiber results in a higher coupling efficiency for direct fusion splicing than tapering using Methods A or B. In contrast, tapering preserves adiabatic mode evolution through a gradual geometric transition from a higher core diameter to a lower one, but does not change the RI of the SMF-28 core. Consequently, although tapering improves MFD matching better than direct fusion splicing, the RI mismatch between the SM1500(4.2) and SMF-28 cores remains largely unchanged. This mismatch limits the achievable coupling efficiency when light propagates from the higher-NA fiber to the lower-NA fiber, leading to higher losses than with direct splicing. This explanation can also support the observation of (ii) better throughput in Method B as compared to Method A. In Method B, direct splicing occurs between pre-tapered SMF-28 and SM1500(4.2), thereby slightly modifying the dopant concentration. We also observe that (iii) reducing the taper diameter of SMF-28 to match the MFD of SM1500 (4.2) did not result in higher transmission, contrary to our expectations. We conducted simulations (RSoft BeamPROP) to further investigate the effects of tapering and splicing on power transmission. Although the simulation results are consistent with some of our observations, the BEAMPROP simulations are constrained by idealized step-index assumptions and do not account for dopant out-diffusion or thermally induced refractive index changes. Consequently, they cannot reproduce the absolute performance of methods that combine tapering and splicing, where such material transformations play a dominant role. Appendix~\ref{app:tapering} presents microscope images of the tapered sections, experimental setup to measure IL, and
simulation details.

 Although tapering did not reduce IL under the conditions investigated here, the results provided important insights into the factors governing coupling between dissimilar fibers, indicating the need for a more rigorous optimization of the taper-splicing sequence that accounts for material-specific thermal behavior and dopant out-diffusion. However, tapering also introduces a practical limitation: the fiber joint becomes mechanically fragile and difficult to handle. We therefore turn to bridging fibers as an alternative IL-reduction strategy, discussed in the following
section.

\subsection{Bridging fibers}
\label{sec:bridging_fibers}

\begin{figure}[htbp]
    \centering
    \begin{subfigure}[b]{0.45\textwidth}
        \centering
        \includegraphics[width=\linewidth]{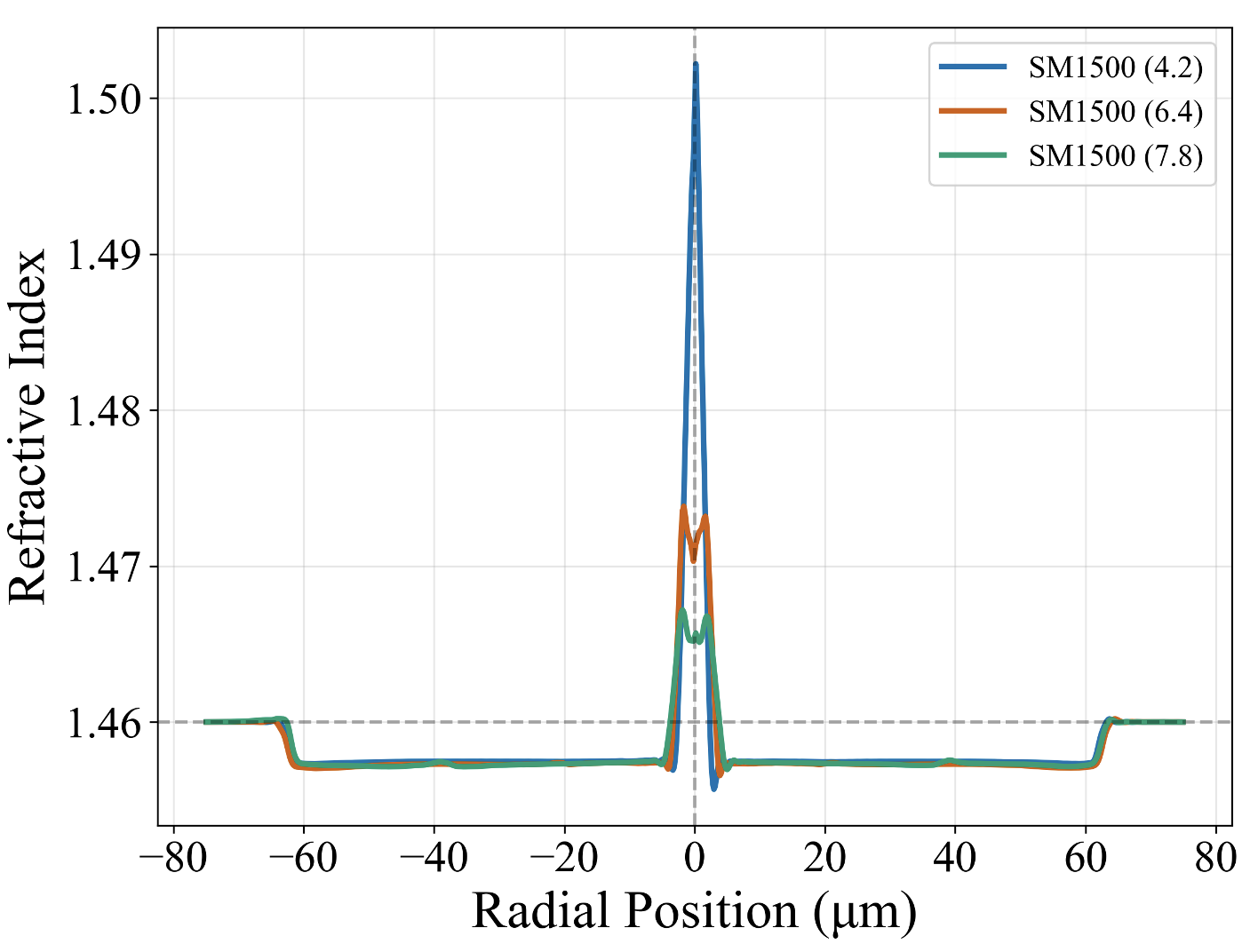}
        \caption{}
    \end{subfigure}
    \hfill
    \begin{subfigure}[b]{0.53\textwidth}
        \centering
        \includegraphics[width=\linewidth]{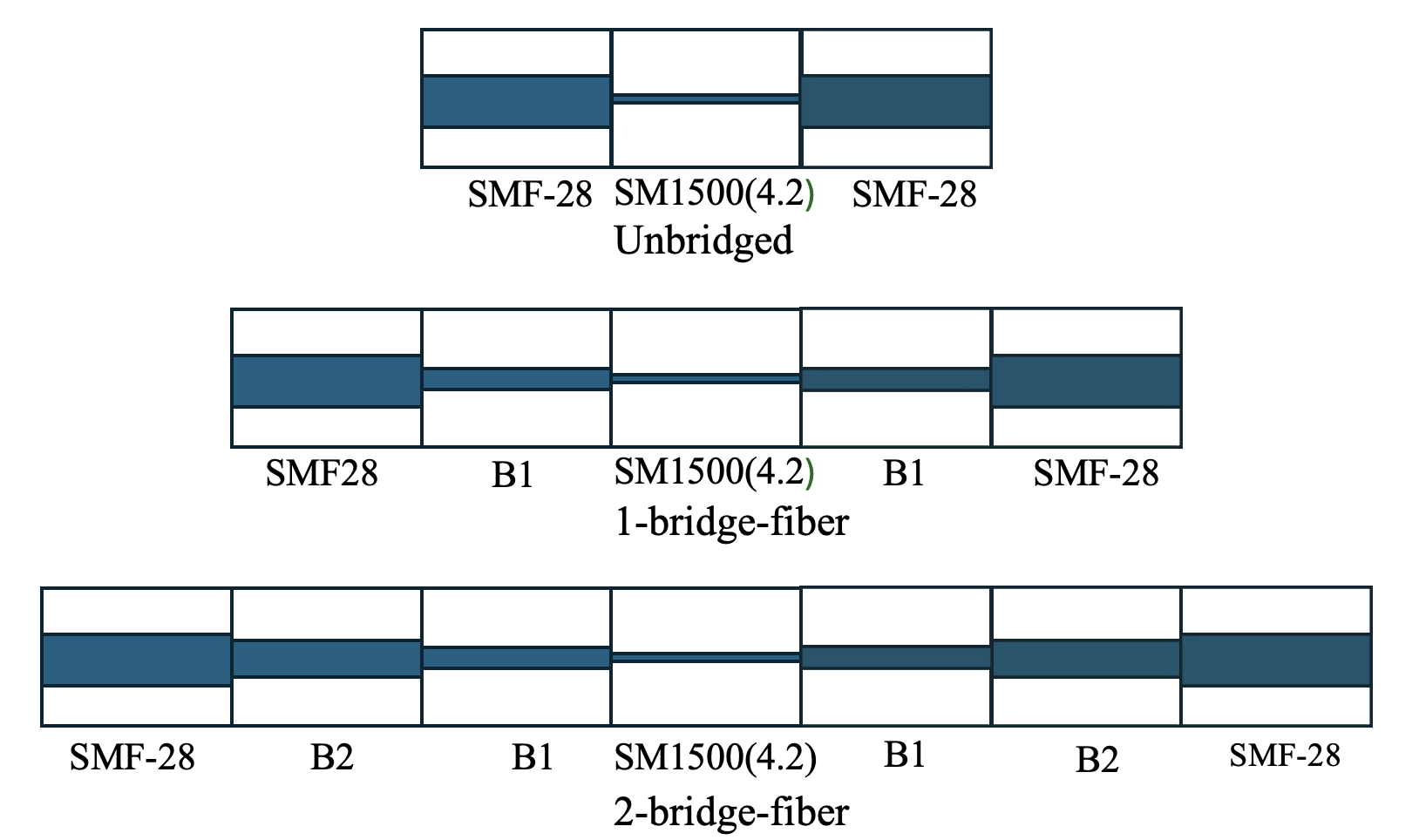}
\caption{}
    \end{subfigure}
    \caption{(a) Refractive indices of core and cladding of SM1500(4.2), B1:SM1500(6.4) and B2:SM1500(7.8) measured using IFA-100 Multiwavelength Optical Fiber Analyzer at \SI{630}{nm}. (b) Schematic illustrating the unbridged and bridged fiber configuration used to connect  SM1500(4.2) (MFD \SI{4.2}{\micro\meter}) with SMF-28. Unbridged: direct fusion splicing between SM1500(4.2) and SMF-28; 1-bridge-fiber: bridged via 1 intermediate/bridging fiber (B1) with MFD of \SI{6.4}{\micro\meter}; 2-bridge-fiber: bridged via 2 intermediate/bridging fibers with MFDs \SI{6.4}{\micro\meter} (B1) and \SI{7.8}{\micro\meter} (B2).}
    \label{fig:bridge_config}
\end{figure}

Using intermediate or “bridging” fibers \cite{Yablon:07}, the splice-loss associated with MFD  mismatch from 
\SI{4.2}{\micro\meter} (SM1500(4.2)) to \SI{10.4}{\micro\meter} (SMF-28) can be minimized. In this technique, the coupling efficiency is improved by stepwise transitioning the MFD using commercially manufactured bridging fibers, SM1500(6.4) (hereafter referred to as B1) and SM1500(7.8) (hereafter referred to as B2), with intermediate MFDs, \SI{6.4}{\micro\meter} and \SI{7.8}{\micro\meter}, respectively. B1 and B2 have the same cladding diameters as that of SM1500(4.2) and SMF-28 (i.e., \SI{125}{\micro\meter}). Fig.~\ref{fig:bridge_config}(a) shows the refractive indices of the core and cladding along the radial position of SM1500(4.2) and the bridging fibers, showing an increasing index contrast for these fibers from larger (\SI{7.8}{\micro\meter}) to smaller (\SI{4.2}{\micro\meter}) MFD.

We used small sections ($<$ \SI{15}{cm}) of SM1500(4.2), B1, and B2 fibers for the configurations shown in Fig.~\ref{fig:bridge_config}(b). The two SMF-28 ends of each configuration were connected to patch cables through FC/APC fiber connectors. Light was launched into each configuration by a BBS (Thorlabs ASE730) via the patch cable. The transmitted power was then  measured by an OSA with a resolution of \SI{50}{pm} (Yokogawa AQ6375B) at \SI{1550}{nm} (refer to Fig.~\ref{fig:Loss_Measurement}(b)). 

We prepared four samples for each of the three configurations (Fig.~\ref{fig:bridge_config}(b)) using the fusion splicer (Sumitomo Electric TYPE-Q101-CA). The angular mismatch between the cleave angles at each splice junction was maintained within  \(0.2^\circ\)-\(1^\circ\). We then measured IL at \SI{1550}{nm} for all 12 samples (4 samples for each of the three configurations). The mean IL and its standard deviation (SD) per configuration are presented in Table \ref{tab:bridge_IL}. Considering the transmission through an SMF-28 as the reference (IL = \SI{0}{dB}), a direct splice between an SMF-28 and an SM1500(4.2), a 1-bridge-fiber connection, and a 2-bridge-fiber connection 
showed average transmission losses of  \SI{4.76}{dB}, \SI{2.55}{dB}, and \SI{1.75}{dB}, respectively. Consequently, as expected, the 2-bridge-fiber connection reduces the mean IL by $\approx$ \SI{3}{dB} compared to the unbridged configuration.

\begin{table}[htbp]
\centering
\caption{Measured IL for configurations between SMF-28 and SM1500 fibers.}
\label{tab:bridge_IL}
\renewcommand{\arraystretch}{1.2}
\setlength{\tabcolsep}{2.8pt}
\begin{tabular}{|p{8.8cm}|c|}
\hline
\centering \textbf{Fiber Bridge Configuration} & \centering \textbf{Measured (mean $\pm$ SD) IL [dB] (@\SI{1550}{nm})} \tabularnewline
\hline
 SMF-28 & reference \\
\hline
  SMF-28 - SM1500(4.2) - SMF-28 & 4.76 $\pm$ 0.24 \\
\hline
SMF-28 - B1 - SM1500(4.2) - B1 - SMF-28 & 2.55 $\pm$ 0.22 \\
\hline
 SMF-28 - B2 - B1 - SM1500(4.2) - B1 - B2 - SMF-28 & 1.75 $\pm$ 0.22 \\
\hline
\end{tabular}
\end{table}
\begin{figure}[H]
\centering
\includegraphics[width=0.7\linewidth]{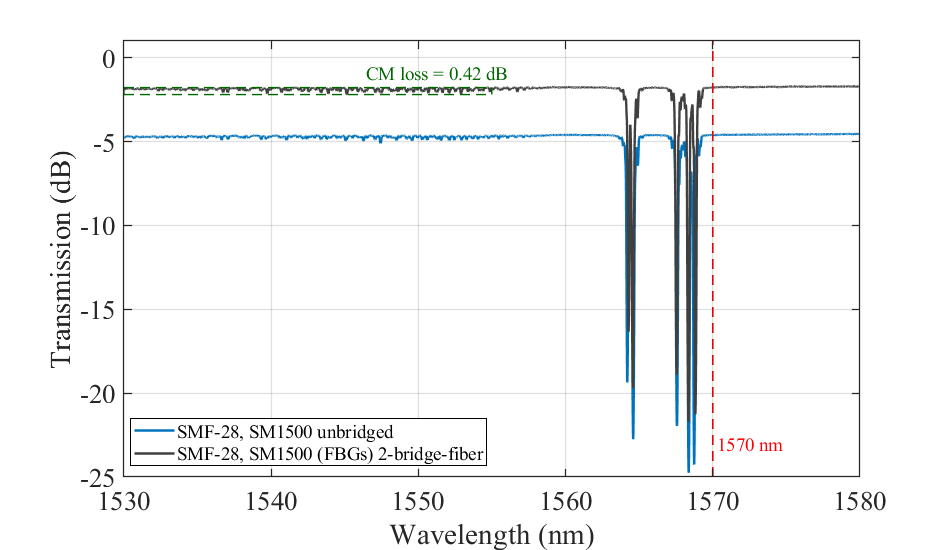}
\caption{Throughput comparison between SMF, SM1500(4.2) unbridged and SMF, SM1500(4.2) 2-bridge-fiber at \SI{1570}{nm}. In the 2-bridge-fiber configuration, IL improves from \SI{4.7}{dB} to \SI{1.77}{dB}.}
\label{fig:FBG_bridge_results}
\end{figure}

Finally, we compared the throughput of the FBG filters inscribed in SM1500(4.2) (Section \ref{SM1500(4.2)}) for the unbridged and 2-bridge-fiber configurations (refer to Fig.~\ref{fig:bridge_config}(b)). As shown in Fig.~\ref{fig:FBG_bridge_results}, the 2-bridge-fiber configuration yields a significant IL improvement of $\approx$ \SI{3}{dB} at \SI{1570}{nm}, reducing the IL from \SI{4.7}{dB} to  $\approx$ \SI{1.77} {dB} and improving the overall transmission from $\approx$ 33.9\% to $\approx$ 66.5\% compared to the unbridged configuration.  

Unlike tapered fibers, the bridging configuration retains the original cladding diameter at each splice junction, making it mechanically more robust and easier to handle. The bare splice sections can be further protected by standard means such as recoating, V-groove support, glass capillary sleeving, or low-shrinkage adhesive encapsulation, all of which are routinely used in fiber device packaging.

\section{Integration into Astrophotonic System}
So far, we have compared the IL and CM losses of filters inscribed by UV- and fs-IR-illumination in several photosensitive fibers, as well as in SMF-28 as a stand-alone platform. To extend this assessment to next-generation miniaturized instrumentation, we now analyze the losses of these filters under complete integration into the astrophotonic system.

\begin{figure}[H]
\centering
\includegraphics[width=\linewidth,height=5cm]{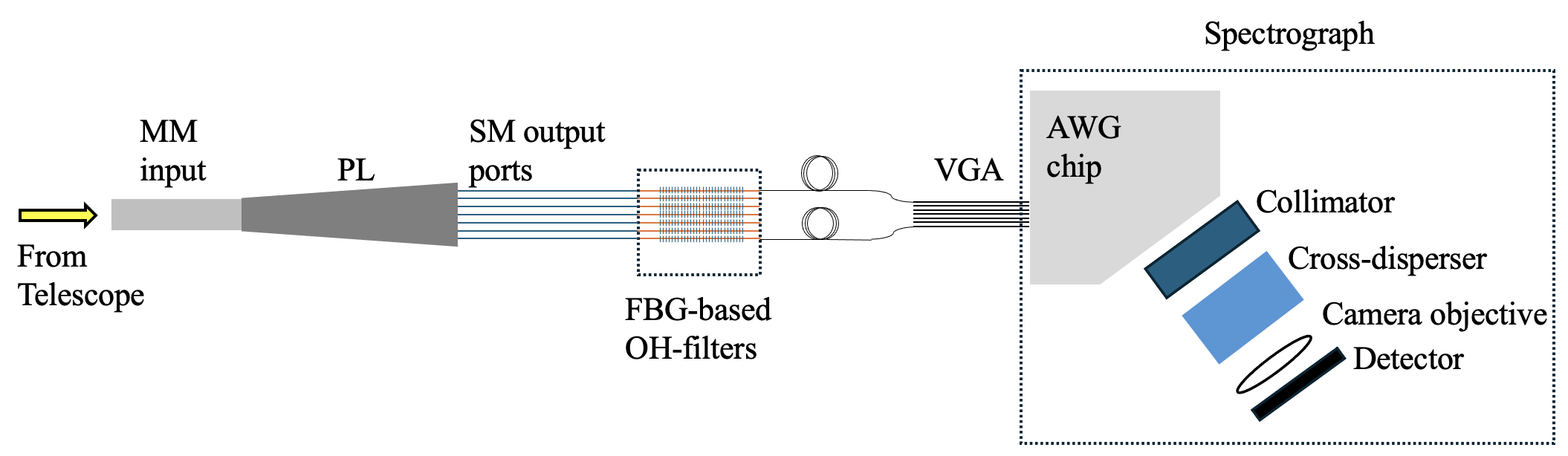}
\caption{Schematic of an astrophotonic spectrograph: PL: Photonic lantern, MM: Multi-mode, SM: Single-mode, FBGs: Fiber Bragg grating OH filters, VGA: V-groove array, AWG: Arrayed waveguide grating.}
\label{fig:schematic_astrophotonic}
\end{figure}

For next-generation extremely large telescopes, a major challenge in conventional astronomical instrumentation, arising from the scaling relation with telescope mirror size, can be overcome by using PLs \cite{Leon-Saval:2005, Leon-Saval:2010}, a novel fiber-based innovation introduced by astrophotonics. A PL efficiently transforms the multi-mode (MM) telescope light into multiple single-mode (SM) outputs, facilitating the integration of OH-filters inscribed in SM fibers downstream in the instrumentation \cite{Joss:2011}. The light filtered by the OH-filters is then dispersed and analyzed by a photonic chip-based spectrograph \cite{Allington_Spectrograph,Joss_Spectrograph}.
A wide range of astrophotonic spectrographs is currently being developed \cite{Gatkine:19}, with the most widely explored on-chip dispersion technique, based on an arrayed waveguide grating (AWG) \cite{Gatkine:17, Stoll:21}, serving as the primary dispersive element. 
Fig.~\ref{fig:schematic_astrophotonic} shows the astrophotonic components, including a PL, FBG OH-filters, and an AWG-based photonic spectrograph.

\subsection{Multi-input fiber-chip interface}
In the integration of the OH-filters with the PIC, a key consideration is efficient fiber-to-chip coupling, as this is the largest loss component in an astrophotonic spectrograph \cite{Gatkine:24, Jovanovic_2023} which arises from coupling between the fiber and the small-MFD of the silicon waveguide of the AWG chip. To reduce this mismatch, ultra-high-NA (UHNA) fibers are typically used as bridge fibers between the fiber device and the photonic chip due to their small MFDs, typically in the range of \SI{3}{\micro\meter} to \SI{6}{\micro\meter}. In this section, we analyze the loss components for each filter set inscribed in SM1500(4.2) and SMF-28, and propose a fiber-to-chip interface system for integrating the filters with PAWS \cite{Hernandez}, an in-house photonic spectrograph.

The key photonic component in PAWS is an AWG chip. The input waveguides are $\approx$ \SI{127}{\micro\meter} apart from each other and have a waveguide MFD of \SI{4.2}{\micro\meter} at \SI{1550}{nm}. A fiber V-groove array (refer Fig.~\ref{fig:schematic_astrophotonic}) with pre-connected reduced-clad UHNA fibers is used at the fiber-to-chip interface.
We propose the use of reduced-clad UHNA fiber, preferably with a cladding diameter of \SI{80}{\micro\meter} at the chip interface, as it
 offers extremely low bend sensitivity and low coupling losses at the fiber-to-chip interface. In addition, a reduced-clad fiber enables greater flexibility in aligning the fiber from the VGA with the corresponding input waveguide of the AWG. 

\begin{figure}[H]
\centering
\includegraphics[width=0.7\linewidth]{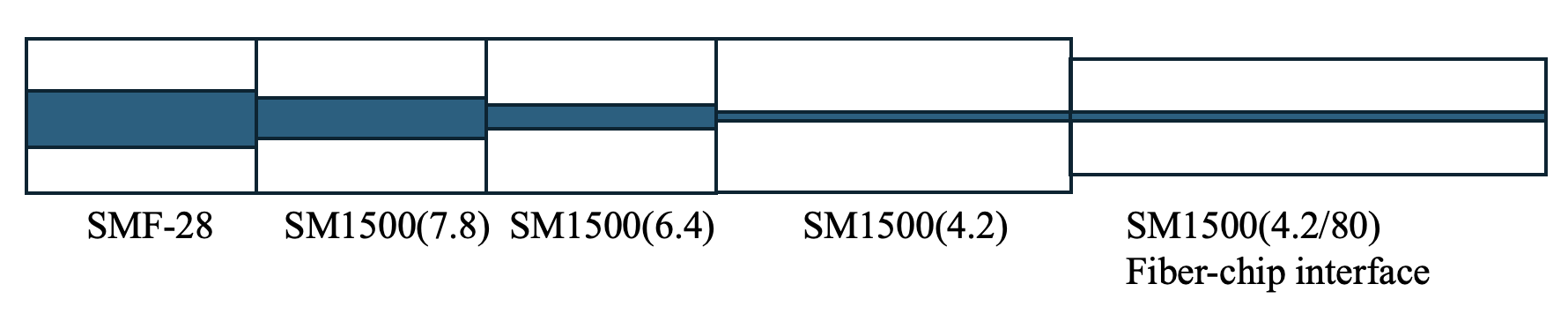}
\caption{Filter integration with astrophotonic spectrograph via reduced-clad UHNA fiber, SM1500(4.2/80) for both UV-inscribed filters in SM1500(4.2) and fs-IR-inscribed filters in SMF-28.}
\label{fig:filter_chip}
\end{figure}

We choose SM1500(4.2/80), with an MFD of $\SI{4.2}{\micro\meter}$ at \SI{1550}{nm} and a cladding diameter of $\SI{80}{\micro\meter}$, as the bridge fiber between the filters and the AWG chip. The UV-inscribed filters in the SM1500(4.2) (standard \SI{125}{\micro\meter} cladding) can directly be spliced to SM1500(4.2/80) at one end. For a maximum angular misalignment of $\theta\leq1^{\circ}$, the maximum splice-loss estimated using Eq.\ref{eq:Mar} is as low as $\SI{0.05}{dB}$. Although the fusion splicing between fibers with dissimilar cladding diameters is often challenging, owing to the excellent core concentricity of the fiber ($\leq \SI{0.5}{\micro\meter}$), low-loss and highly reliable fusion splicing can be achieved between SM1500(4.2) and SM1500(4.2/80) using precision splicers, e.g., Vytran glass processing system. For the complete integration of these filters in the astrophotonic chain of devices as illustrated in Fig.~\ref{fig:schematic_astrophotonic}, the SM1500(4.2) must be connected to the SM end of the PL to couple light from the telescope. Therefore, one end of the SM1500(4.2) must be bridged to one of the SMF-28 fiber ports of the PL. As only one side bridging of SM1500(4.2) is needed as shown in Fig.~\ref{fig:filter_chip}, the SM1500(4.2) bridged configuration will now incur only half of the splice-loss, i.e., IL (considering a negligible fiber attenuation) of $\approx$ \SI{0.88}{dB} as compared to both-sided bridging as shown in Fig.~\ref{fig:bridge_config}(b), with IL of $\approx$ $\SI{1.77}{dB}$ (see Fig.~\ref{fig:FBG_bridge_results}). Therefore, we can infer that UV-inscribed filters in SM1500(4.2) will incur an IL $<$$\SI{1}{dB}$ when these FBG-based OH-filters are introduced between the output of the PL and the input of the VGA. 

\begin{figure}[H]
\centering
\includegraphics[width=0.5\linewidth]{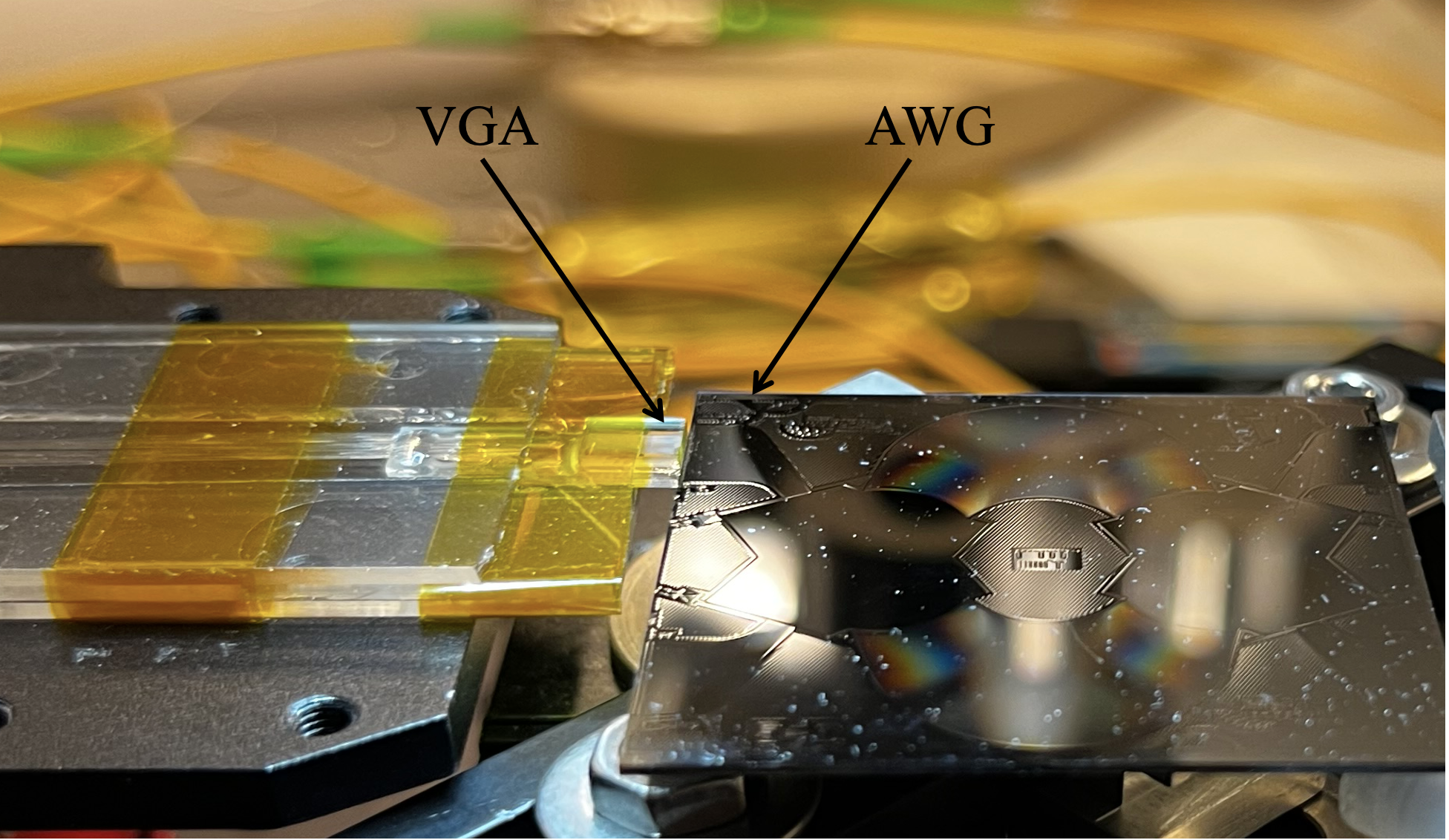}
\caption{Laboratory setup for a fiber V-groove array hosting 16 fibers connected to the multiple input waveguides of an AWG chip.}
\label{fig:VGA_AWG}
\end{figure}

When we consider fs-IR-inscribed filters in SMF-28 for such an integration to an astrophotonic spectrograph, to mitigate the large mismatch of MFD between an SMF-28 and the AWG, a similar bridging option that has been used for filters in SM1500(4.2) can be implemented in these filters to gradually reduce the splice-loss, or in other words, the IL. In this case, a similar IL $<$$\SI{1}{dB}$ is expected. 
Both SM1500(4.2) and SMF-28, which have a cladding diameter of \SI{125}{\micro\meter}, are spliced/connected to a reduced-clad fiber with a cladding diameter of \SI{80}{\micro\meter}. It is not straightforward to assess how incorporating a reduced-clad fiber at the interface will affect CM losses for both fiber platforms, warranting further study. We are currently setting up a laboratory demonstration (to be presented in future communications) of a multi-input AWG using a custom-made 16-channel V-groove assembly (OZOptics VGA) hosting 16 arrays of fibers, SM1500 (6.4/80) (Fibercore) (Fig.~\ref{fig:VGA_AWG}).

\subsection{Multi-input fiber-chip interface: reduction of IL using reduced-clad fiber}
Based on published \cite{Fibercore} splice-loss data from several reduced-clad fiber configurations of the SM1500 series that bridge with a fiber with MFD of $\sim$ \SI{4.2}{\micro\meter}, we can further minimize the IL for both fiber platforms containing the OH filters. Table \ref{tab:splice_loss_chart} shows three such bridging fiber configurations.

\begin{table}[htbp]
\centering
\caption{Measured splice-loss configurations between SMF-28 and SM1500 fibers \cite{Fibercore}.}
\label{tab:splice_loss_chart}
\renewcommand{\arraystretch}{1.2}
\begin{tabular}{|c|c|c|c|c|c|}
\hline
\textbf{Type} & \textbf{Fiber 1} & \textbf{Fiber 2} & \textbf{Fiber 3} & \textbf{Fiber 4} & \textbf{Average Splice-loss} \\

\hline
\textbf{1} & \textbf{SMF-28} & \textbf{SM1500 (7.8/80)}& \textbf{SM1500 (4.2/80)}& -- & \textbf{0.44\,dB} \\
\hline
2 & SMF-28 & SM1500 (7.8/80)& SM1500 (4.2/50)& -- & 0.94\,dB \\
\hline
3 & SMF-28 & SM1500 (7.8/80)& SM1500 (6.4/80) & SM1500 (4.2/50) & 0.55\,dB \\
\hline
\end{tabular}
\end{table}

\begin{figure}[H]
\centering
\includegraphics[width=0.65\linewidth]{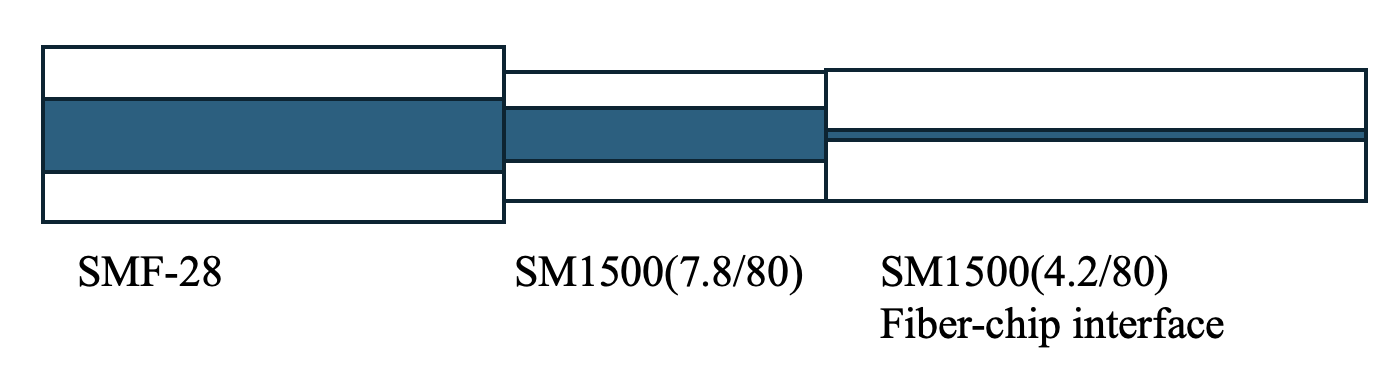}
\caption{Filter integration with astrophotonic spectrograph, with reduced IL for both filter types, inscribed with UV-illumination in SM1500(4.2/80) and with fs-IR-illumination in SMF-28.}
\label{fig:propose}
\end{figure}

For our application, configuration type 1 seems to be the best option for an average splice-loss of \SI{0.44}{dB}. Type 1 configuration connects fiber 1, SMF-28 to fiber 2, a reduced-clad fiber, SM1500(7.8/80) of cladding diameter \SI{80}{\micro\meter} and MFD of \SI{7.8}{\micro\meter}  and then finally to fiber 3, SM1500(4.2/80) of cladding diameter \SI{80}{\micro\meter} and MFD of \SI{4.2}{\micro\meter}. SM1500(4.2/80) can then be connected to the input waveguide of the AWG chip in PAWS. Based on the bridging configuration type 1, we propose a bridging configuration for both the filter types as shown in  Fig.~\ref{fig:propose}. When fs-IR-illumination is used, filters can be directly inscribed into SMF-28 or into any downstream bridge fiber. With UV-illumination, as the UHNA fibers from the SM1500 series also have very high germania concentration and therefore high photosensitivity, filters can be directly inscribed within the SM1500(4.2/80) fiber instead of the SM1500(4.2). In this proposed configuration, not only will the IL of our FBG-based filters be reduced to $\approx$ \SI{0.5}{dB} from \SI{1}{dB} of the configuration as discussed in the previous section, we will also be able to reduce the number of splice junctions from 4 (Fig.~\ref{fig:filter_chip}) to 2 (Fig.~\ref{fig:propose}), i.e., by 50\%. This decrease in splice junctions will be significant when we consider the overall system with multiple PLs. For a single PL with 19 SM output ports, each with sets of OH filters, we will minimize the number of splice junctions from 76 to 38. However, due to the reduced fiber cladding diameter, we cannot comment on the CM losses at this point without conducting further studies. % for 19 sets of OH filters. 
To implement this proposed configuration, we plan to include the following in our future studies: 
(i) experimental validation of the splice-loss, i.e., IL, for the proposed configuration (it should be noted that the splice-loss values in Table~\ref{tab:splice_loss_chart} are drawn from a manufacturer's technical bulletin~\cite{Fibercore}) using Vytran glass processing system,
(ii) assessing CM losses in FBGs inscribed in a reduced-clad fiber SM1500(4.2/80) with UV-illumination,
(iii) improving fs-IR-illumination combining aperture control and minimum  CM loss, and finally, 
(iv) evaluating the UV and fs-IR filter types by conducting an overall filter loss assessment in a multi-input fiber connection to AWG of PAWS via a custom-made VGA fiber assembly. Based on studies limited to a few filter lines, the goal is to extend this to all $H$ band filter lines to develop a complete astrophotonic instrument with OH-filtering capabilities for ground-based telescopes. 

\section{Conclusions}
We have presented a comparative study of insertion and cladding mode losses in five FBG-based OH-filter lines in different fibers under UV- and fs-IR-illumination. Our study demonstrates a clear trade-off between the two approaches when assessed as stand-alone: under the UV inscription technique, the photosensitive fiber SM1500(4.2), with high-NA of $\approx$ 0.3 and small MFD of $\approx$ \SI{4.2}{\micro\meter} at \SI{1550}{nm}, demonstrates the best performance in terms of CM losses, while SMF-28 exhibits the lowest IL when filters are inscribed in these fibers using fs-IR. As FBGs are fabricated in SM fibers, to benefit from these filters for astronomical applications, in the upstream, the filters must be integrated into a PL that collects light from the telescope via MM fibers and then distributes it to several SM output ports. 
The filters in SM1500(4.2), when connected to the SM output ports (SMF-28) of a PL, incur significant IL, primarily due to high splice-losses at the fusion splice junction caused by the large MFD mismatch between SM1500(4.2) and SMF-28. We investigated two approaches to mitigate the high IL in SM1500(4.2) using tapering and bridge-fiber options and demonstrated that the bridge-fiber option offers superior performance, achieving an IL improvement of $\approx$ \SI{3}{dB}. 
In contrast, fs-IR-inscribed filters exhibit superior IL because they are already inscribed within SMF-28. However, the implemented aperture shaping technique for controlling the spectral profiles of the fs-IR-written filters in SMF-28 introduces higher CM losses than the UV-written filters in SM1500(4.2).
We further aim to develop a complete astrophotonic instrument for ground-based NIR observations. We
presented a preliminary evaluation of both filter types when integrated with an in-house-developed AWG-based photonic spectrograph. At the chip interface, to maximize fiber-to-chip coupling efficiency, SM1500(4.2/80), a reduced-clad fiber (cladding diameter of \SI{80}{\micro\meter}) with an MFD of \SI{4.2}{\micro\meter}, is used. To implement the astrophotonic instrument with OH-line filtering capabilities, filters must be connected to SM1500(4.2/80) at the downstream end, resulting in an IL < \SI{1}{dB} for both filter types. It is important to note that when both the UV- and fs-IR-inscribed filters are connected to the PL upstream and the AWG chip downstream, the distinction between the two filter types with respect to IL becomes less significant. At the same time, CM losses would need further investigation to determine whether the reduced-clad fiber at the fiber-to-chip interface alters them.
Based on the present results, we conclude that although stand-alone UV-inscribed filters in SM1500(4.2) appear disadvantaged, both UV-inscribed filters in SM1500(4.2) and fs-IR-inscribed filters in SMF-28 remain viable platforms for OH-suppression filters when system-level performance within the astrophotonic architecture is considered.

Finally, we have proposed a compact bridge-fiber configuration employing reduced-cladding-diameter fiber, applicable to both UV- and fs-IR-inscribed filters for astrophotonic spectrographs, which can further improve the IL  (IL$<$\SI{0.5}{dB}) while reducing the number of splice junctions between the bridging fibers by 50\%. Future studies will focus on assessing these configurations when directly interfaced between a PL and a photonic chip, enabling selection of the appropriate fiber platform and inscription method for OH-suppression in future compact astrophotonic spectrographs for ground-based NIR telescopes.

\appendix   
\section{Adiabatic Tapering: Microscope images, IL measuring setup, and simulation}
\label{app:tapering}
 
This appendix provides supplementary details for Section~\ref{IL_AdbTap},
comprising microscope images of tapered splice samples, the
experimental setup used for transmission loss measurements, and the
beam-propagation simulation used to interpret the results.

\subsection{Microscope images of tapered splice samples}

\begin{figure}[htbp]
\centering
\includegraphics[width=0.8\linewidth]{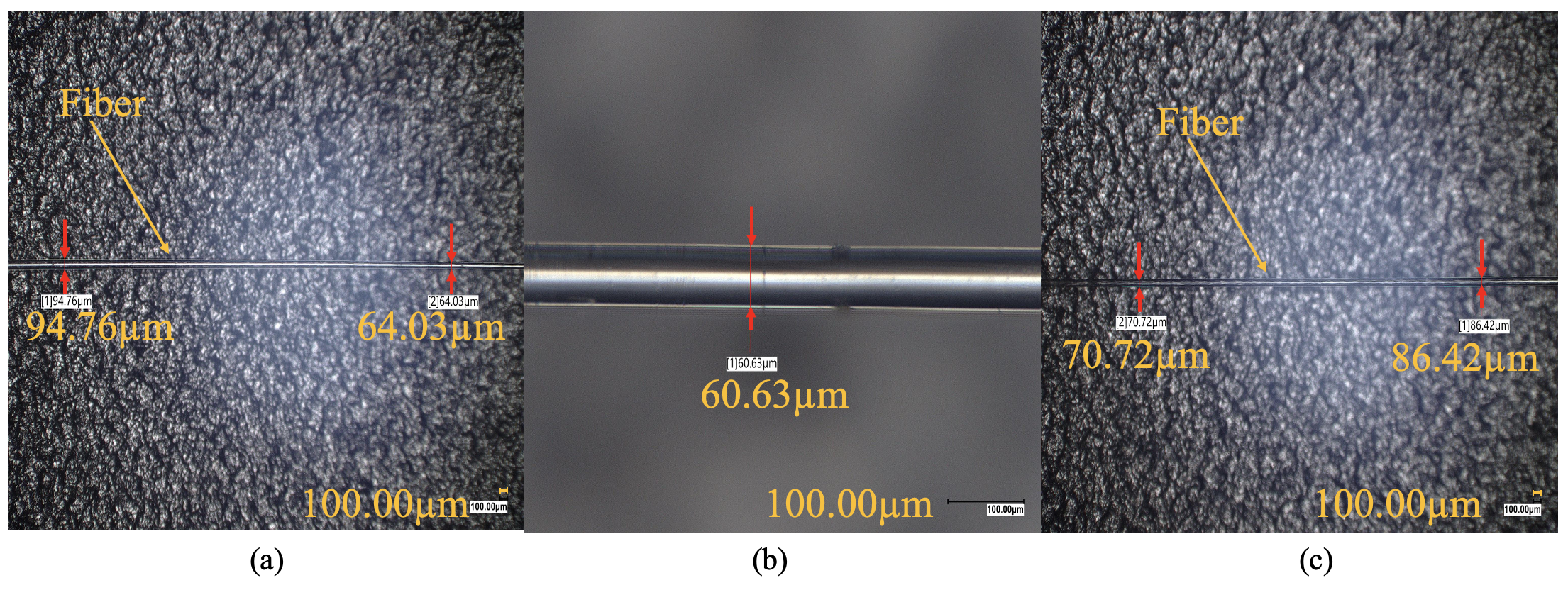}
\caption{Microscope images of Method~A tapered section (refer to
Fig.~\ref{fig:schematic}(b)). Three micrographs stitched to show the
full transition: (a)~50$\times$ magnification of the SMF-28 downtaper,
(b)~500$\times$ magnification of the splice region at a cladding diameter
of $\sim$\SI{60}{\micro\meter}, (c)~50$\times$ magnification of the
SM1500(4.2) uptaper. Post-splice tapering makes the junction interface
indistinguishable in (b).}
\label{fig:MethodA}
\end{figure}

\begin{figure}[htbp]
\centering
\includegraphics[width=0.3\linewidth]{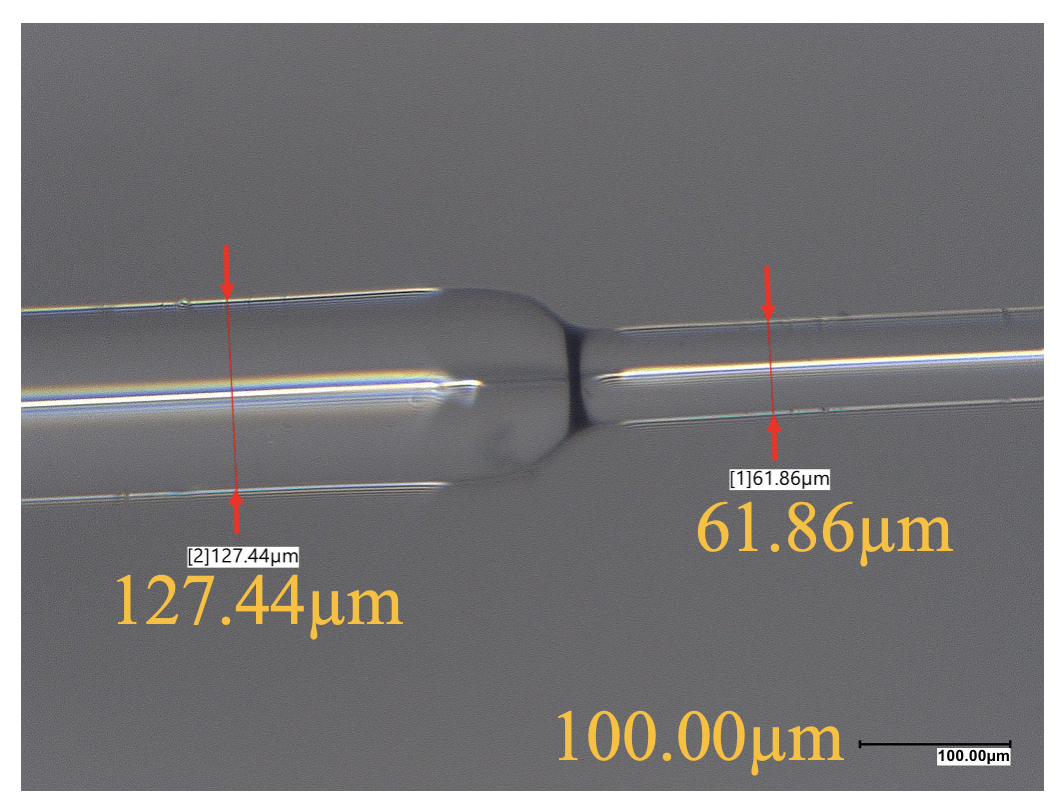}
\caption{Microscope image of Method~B tapered section
(refer to Fig.~\ref{fig:schematic}(b)): 500$\times$ magnified view of
the fusion-splice interface, with SM1500(4.2) on the left and pre-tapered
SMF-28 on the right at a cladding diameter of $\sim$\SI{60}{\micro\meter}.
The splice junction is clearly visible.}
\label{fig:MethodB}
\end{figure}

Microscope images of representative samples from Method~A (Fig.~\ref{fig:MethodA}) and Method~B (Fig.~\ref{fig:MethodB}),
both with a tapered cladding diameter of $\sim$\SI{60}{\micro\meter},
were captured using a Keyence VH-Z50L microscope. The annotations are
rewritten in yellow adjacent to the original microscope labels for
readability. 
 
\subsection{IL measurement setup}

\begin{figure}[htbp]
\centering
\includegraphics[width=\linewidth]{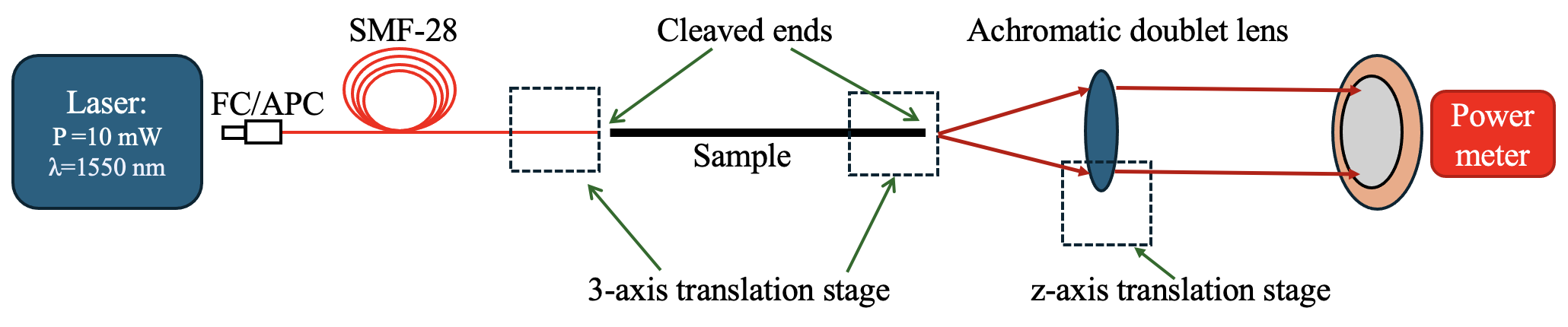}
\caption{Experimental setup for measuring transmission power across
different splicing configurations. Sample~A: tapered by Method~A;
Sample~B: tapered by Method~B (refer to Fig.~\ref{fig:schematic}(b)).}
\label{fig:exp_setup}
\end{figure}

Fig.~\ref{fig:exp_setup} shows the experimental setup used to measure the
transmitted power across the three splicing configurations (direct fusion
splice, Method~A, and Method~B). A tunable laser source ($\lambda =
\SI{1550}{nm}$, $P = \SI{10}{mW}$) is connected to an SMF-28 input fiber
via an FC/APC connector; its cleaved end is mounted on a 3-axis translation
stage for precise alignment to each sample. At the output end, a second
3-axis translation stage holds an achromatic doublet lens that focuses
the transmitted beam onto a power meter, maximizing collection efficiency.
 
% -----------------------------------------------------------
\subsection{Simulation}
The simulations were performed under the following assumptions:

\begin{enumerate}[topsep=3pt, itemsep=3pt, parsep=1pt, partopsep=0pt,label=\roman*)]
    
    \item The core diameter, $d_i$ and RI values of the core, $n_{core}$, and the cladding, $n_{clad}$ were based on measurements obtained from our RI profilometer (IFA-100 Multiwavelength Optical Fiber Analyzer). We used these values for SMF-28: $d_i = \SI{8.6}{\micro\meter}$, $n_{core}=1.46366$, $n_{clad} =1.45776$, and for SM1500(4.2): $d_i = \SI{3.2}{\micro\meter}$, $n_{core}=1.49787$, $n_{clad} =1.45718$. The simulation does not account for the thermal dopant out-diffusion inherent in glass-fusion processes.
    
    \item We modeled the fiber using a step-index RI profile. 
    
    \item For single-mode fibers, since the core diameter is much smaller than the cladding diameter, we modeled the cladding to be infinitely large to reduce computational time.
    
    \item In our tapering processes, as the SMF-28 fiber is always tapered down due to its large diameter, we define the tapered diameter ratio ($D_t$) as the ratio of the initial core diameter ($d_i$) to the final core diameter ($d_f$). 
\end{enumerate}

\begin{figure}[h]

\begin{subfigure}{0.49\textwidth}
\includegraphics[width=1.0\linewidth]{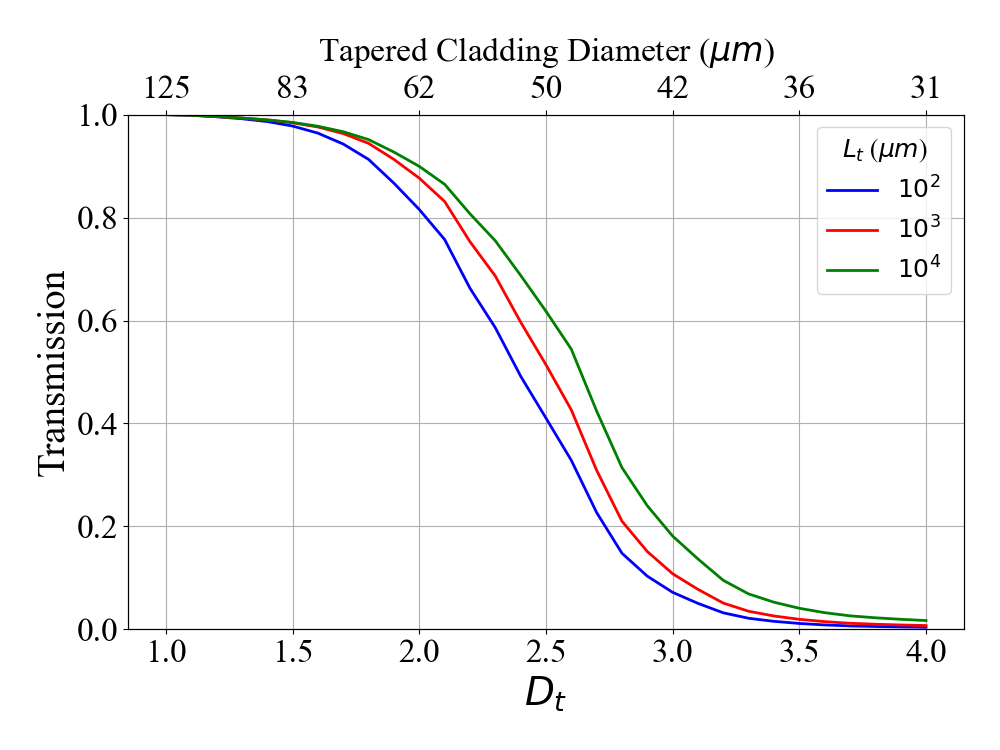} 
\caption{}
\label{fig:smf28_paper}
\end{subfigure}
\hfill
\begin{subfigure}{0.49\textwidth}
\includegraphics[width=1.0\linewidth]{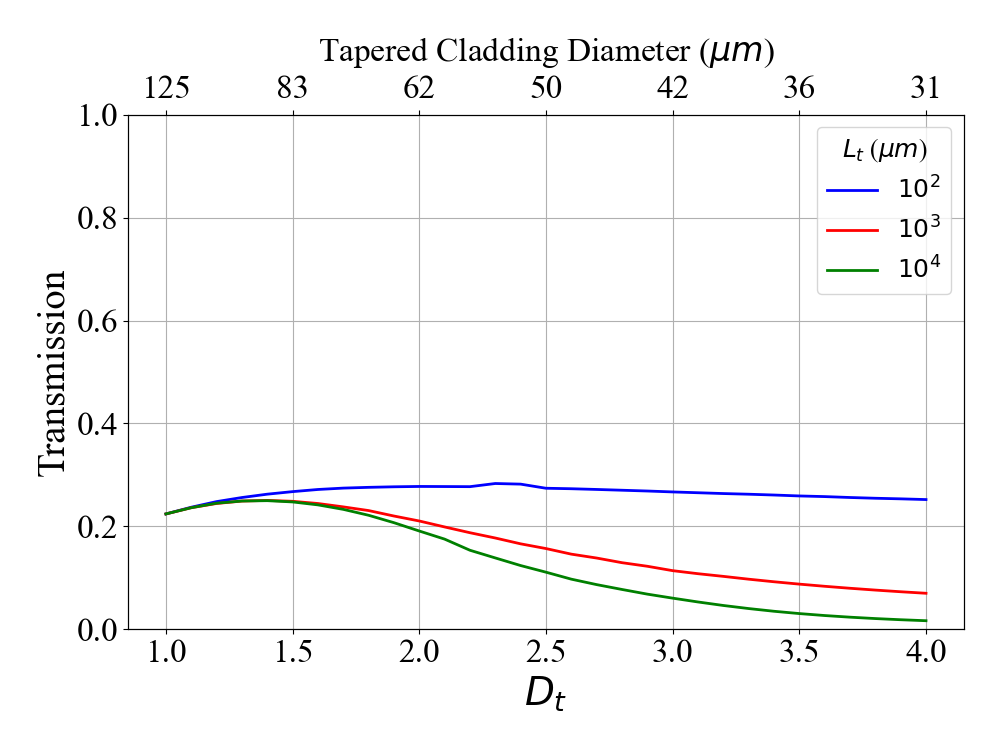}
\caption{}
\label{fig:smf28_sm1500_paper}
\end{subfigure}

\caption{Simulated normalized transmission as a function of tapered diameter ($D_t$) for three different taper lengths ($L_t$). (a) Response for a standalone SMF-28 fiber. (b) Response of an SMF-28/SM1500(4.2) system, demonstrating the combined transmission resulting from both tapering and splicing.}
\label{fig:smf28_a_b}
\end{figure}

Under the above assumptions, the BeamPROP module in RSoft was used to simulate the propagation of the fundamental mode of the SMF-28 fiber at $\lambda = \SI{1550}{nm}$. To evaluate the modal evolution, the normalized transmission at the output was calculated as a function of $D_t$ across three distinct taper lengths ($L_t$): \SI{e2}{\micro\meter}, \SI{e3}{\micro\meter} and \SI{e4}{\micro\meter}. This range of short, intermediate, and long transition regions was selected to illustrate the performance trends and the transition toward adiabatic conditions. The transmission characteristics of the tapered SMF-28 fiber is shown in Fig.~\ref{fig:smf28_a_b} (a). For a fixed taper length, the transmission follows an inverted S-shaped curve, which can be described mathematically by a sigmoidal decay function.  
When the diameter is not tapered ($D_t \sim 1$), the transmission is $\sim$ 1, indicating that the fundamental mode is well confined in the core of the fiber. 
The transmission drops sharply in the range of $2 < D_t < 3$, as the fundamental mode begins to couple into cladding or radiation modes, essentially increasing the adiabatic loss (the adiabatic loss can be reduced by increasing the taper length). At $D_t > 3.5$, the curve tends to flatten out, making the transmission $\sim$ 0, implying that the adiabaticity condition is no longer satisfied.

The introduction of SM1500(4.2) fiber significantly alters the transmission profile due to the combined effects of tapering and splicing, as shown in Fig.~\ref{fig:smf28_a_b} (b). At $D_t = 1$, a $78 \%$ power loss (transmission = 0.22) is observed due to modal mismatch at the fiber junction.  
When compared with Fig.~\ref{fig:smf28_a_b} (a), in the combination of tapering and splicing, the curves exhibit a positive skew. Tapering slightly improves the transmission relative to the baseline splice results: intermediate and long tapers have a transmission peak at 0.25 ($D_t \approx 1.4$), while the short taper ($L_t = \SI{100}{\micro\meter}$) shows a $27 \%$ improvement at $D_t = 2$. This optimal $D_t$ closely matches the MFD ratio ($\sim 2$) between the two fibers used in this experiment, suggesting that tapering the SMF-28 can partially compensate for the initial modal mismatch due to the splice alone. Furthermore, as the experimental taper diameters in Fig.~\ref{fig:Results} fall within the $1.6 < D_t < 2.1$, the simulation results in Fig.~\ref{fig:smf28_a_b} (b) support the experimental observations in Fig.~\ref{fig:Results}: the transmission obtained after splicing alone exceeds that obtained after the combined splicing and tapering processes. However, the overall loss of the experimental system was significantly higher, at $87\%$, than in the simulation, at $78\%$. This additional loss in the experiment can be attributed to several factors: connector losses between the fiber and the laser, finite attenuation of the fibers used in the experiment, Fresnel losses at the fiber-air-fiber interface, coupling losses due to misalignment, and manufacturing imperfections in both the tapering and splicing processes. 
Simulation results indicate that transmission improves as $D_t$ decreases or tapered cladding diameter increases; however, since the experimental taper length of \SI{4}{cm} exceeds the upper bound of the simulated range ($L_t$ = \SI{e4}{\micro\meter}), the simulation provides a qualitative basis for interpreting the trend rather than a direct quantitative prediction at the experimental length scale. However, no clear trend was observed in the experimental data, likely due to non-idealities in both the manufacturing Methods A and B. Furthermore, the high photosensitivity of the SM1500(4.2) fiber, which contains five times the germania concentration of standard SMF-28, likely leads to complex dopant out-diffusion during the fusion process. Such material transformations are not captured in idealized step-index simulations.

\subsection*{Funding}
This work is supported by DFG (Deutsche Forschungsgemeinschaft project no.~455425131) and EFRE StaF - Verbund project PICS4SENS (project no.~86000879).

\subsection*{Disclosures}
The authors declare that there are no financial interests, commercial affiliations, or other potential conflicts of interest that could have influenced the objectivity of this research or the writing of this paper.

\subsection* {Code, Data, and Materials Availability} Data underlying the results presented in this paper are not publicly available at this time but may be obtained from the authors upon reasonable request.

\subsection* {Acknowledgments}
The authors thank Fibercore Limited and Laser Components Germany GmbH for providing sample bridging fibers and valuable suggestions, and Coherent for providing the information about cladding mode suppressed fibers. P. Choochalerm acknowledges support from the Science and Technology Development Fund (ST68), Thailand, for a visiting scientist fellowship at the Leibniz Institute for Astrophysics Potsdam (AIP). The authors used AI tools to clean up the language and grammar during the preparation of the manuscript. All scientific content, experimental data, analysis, interpretation, and conclusions are entirely the authors' own work. 

%%%%% References %%%%%

\bibliography{report}   
\bibliographystyle{spiejour}

\vspace{2ex}\noindent\textbf{Aashia Rahman} is a Senior Scientist at the Leibniz Institute for Astrophysics Potsdam (AIP), with more than 20 years of combined research and industry experience. She received her PhD in Instrumentation and Applied Physics from the Indian Institute of Science (IISc), Bangalore, in 2009. Her research interests include fiber optics in astrophotonics, fiber-to-chip coupling, and the development of specialized filters for ground-based astronomical instrumentation.

\vspace{1ex}
\vspace{2ex}\noindent\textbf{Ria Krämer} is a member of the Ultrafast Optics Group at the Friedrich Schiller University Jena. Her work concentrates on ultrashort pulse inscription of FBG. The application ranges from high power robust reflectors for fiber laser systems, dispersion tailored chirped fiber Bragg gratings for e.g. single photon generation to adapted spectral filtering for astrophotonics.

\vspace{1ex}
\vspace{2ex}\noindent\textbf{Abani Shankar Nayak} is currently a scientific staff member at the Leibniz Institute for Astrophysics Potsdam (AIP). He is working on the development, production, and characterization of photonic lanterns, which are essential for future astronomical instrumentation and quantum communication. Dr. Nayak completed his PhD in 2022, focusing on pupil remappers and beam combiners for long-baseline interferometry. Following his PhD, he pursued postdoctoral research at the Institute of Applied Physics in Jena, working on thin-film lithium niobate for quantum technology applications. His primary research interests include experimental optics, astronomical instrumentation, integrated photonics, and experimental quantum optics.

\vspace{1ex}
\vspace{2ex}\noindent\textbf{Julius Göhring} is a Master's student in Astrophysics at the University of Potsdam (graduated April 2026) and a Student Assistant at the Leibniz-Institut für Astrophysik Potsdam (AIP) in the Astrophotonics group (innoFSPEC). His work focuses on the fabrication and characterization of photonic lanterns for high-precision spectroscopy and their integration into the MARCOT telescope concept. He has presented this work at international conferences including EPRV 6 (Porto) and the Photonic E-MARCOT Conference (Granada). Alongside his photonics research, he investigates planetary engulfment and its effects on stellar magnetic activity using wide binary star systems.

\vspace{1ex}
\vspace{2ex}\noindent\textbf{P. Choochalerm} is a Research Engineer specialising in fiber optic technologies for astronomical instrumentation. Her work focuses on the development and characterization of optical fiber components for fiber-fed spectrographs.

\vspace{1ex}
\vspace{2ex}\noindent\textbf{Anna Maria Weiß} is a 4th semester physics bachelor student at University of Potsdam. Since more than one year she is a student assistant working in the Astrophotonics group. Her research interests cover telescope instrumentations as FBGs and Frequency Combs, as well as exoplanet characterisation.

\vspace{1ex}
\vspace{2ex}\noindent\textbf{Samuel Lutz Döpfner} is a graduate student of physics with a specialization in optics at the Friedrich Schiller University Jena and a student assistant in the Ultrafast Optics Group under Prof. Nolte at the Institute of Applied Physics. He received his Bachelor's degree in physics on the topic of fs-IR-inscribed fiber Bragg gratings in 2023 in the same group and is currently researching dispersion characteristics of chirped fiber Bragg gratings for his Master's thesis.

\vspace{1ex}
\vspace{2ex}\noindent\textbf{Tim Schleifer} Not Available.

\vspace{1ex}
\vspace{2ex}\noindent\textbf{Kalaga Madhav} is the Head of R \& D Astrophotonics(innoFSPEC), at the Leibniz Institute for Astrophysics Potsdam (AIP). His group develops photonic technologies for astronomical instrumentation, including integrated spectrographs, OH-suppression filters, photonic lanterns, beam combiners, and frequency combs for precision astronomy.

\vspace{1ex}
\vspace{2ex}\noindent\textbf{Martin M. Roth} Not Available.

\vspace{1ex}
\vspace{2ex}\noindent\textbf{Stefan Nolte} Not Available.

\listoffigures
\listoftables

\end{spacing}
\end{document}